\documentclass[useAMS]{mn2e} 
 
\usepackage{graphicx} 
\usepackage{txfonts} 
\newcommand\aj{AJ} 
\newcommand\apj{ApJ}

\newcommand\aap{A\&A} 
\newcommand\mnras{MNRAS} 
\newcommand\apjl{ApJ} 
\newcommand\pasp{PASP} 
\newcommand\nat{Nature}

\newcommand\aapr{ARA\&A}
 
\title[]{Multiple stellar populations in Magellanic Cloud clusters. IV. The double main sequence of the young cluster NGC\,1755}  
\author[A.\,P.\, Milone et al.] 
{A.\,P.\,Milone$^{1}$, A.\,F.\,Marino$^{1}$, F.\,D'Antona$^{2}$, L.\,R.\,Bedin$^{3}$, G.\,S.\,Da Costa$^{1}$, \newauthor
H.\,Jerjen$^{1}$, A.\,D.\,Mackey$^{1}$\,
\\ 
$^{1}$Research School of Astronomy \& Astrophysics, Australian National University, Canberra, ACT 2611, Australia \\
$^{2}$Istituto Nazionale di Astrofisica - Osservatorio Astronomico di Roma, Via Frascati 33, I-00040 Monteporzio Catone, Roma, Italy\\
$^{3}$Istituto Nazionale di Astrofisica - Osservatorio Astronomico di Padova, Vicolo dell'Osservatorio 5, Padova, IT-35122\\
} 
\begin{document} 
\date{Accepted 2016 March 10. Received 2016 March 8; in original form 2016 January 7} 
%\date{Draft Version Mar, 8, 2016} 
 
\pagerange{\pageref{firstpage}--\pageref{lastpage}} \pubyear{2016} 
 
\maketitle 
\label{firstpage} 
 
\begin{abstract}  

Nearly all the star clusters with ages of $\sim$1-2 Gyr in both Magellanic Clouds exhibit an extended main-sequence turn off (eMSTO) whose origin is under debate. The main scenarios suggest that the eMSTO could be either due to multiple generations of stars with different ages or to coeval stellar populations with different rotation rates.  
   
 In this paper we use {\it Hubble-Space-Telescope} images to investigate the $\sim$80-Myr old cluster NGC\,1755 in the LMC. We find that the MS is split with the blue and the red MS hosting about the 25\% and the 75\% of the total number of MS stars, respectively. Moreover, the MSTO of NGC\,1755 is broadened in close analogy with what is observed in the $\sim$300-Myr-old NGC\,1856 and in most intermediate-age Magellanic-Cloud clusters.
We demonstrate that both the split MS and the eMSTO are not due to photometric errors, field-stars contamination, differential reddening, or non-interacting binaries. These findings make NGC\,1755 the youngest cluster with an eMSTO.

We compare the observed CMD with isochrones and conclude that observations are not consistent with stellar populations with difference in age, helium, or metallicity only. On the contrary, the split MS is well reproduced by two stellar populations with different rotation,  although the fit between the observed eMSTO and models with different rotation is not fully satisfactory. 

We speculate whether all stars in NGC\,1755 were born rapidly rotating, and a fraction has slowed down on a rapid timescale, or the dichotomy in rotation rate was present already at star formation. We discuss the implication of these findings on the interpretation of eMSTO in young and intermediate-age clusters.
\end{abstract} 
 
 
\section{Introduction}\label{sec:intro} 
 Several studies have revealed that most Globular Clusters (GCs) with ages of $\sim$1-2 Gyr in both Magellanic Clouds (MCs) exhibit either a bimodal or extended main-sequence turn-off (hereafter eMSTO, Bertelli et al.\,2003; Mackey \& Broby Nielsen 2007; Glatt et al.\,2009; Milone et al.\,2009;  Correnti et al.\,2014; Goudfrooij et al.\,2014).   
 This feature of the color-magnitude diagram (CMD) has suggested that eMSTO GCs have experienced a prolonged star-formation history with a duration of $\sim$100-500 Myrs (e.g.\,Mackey et al.\,2008; Rubele et al.\,2013; Goudfrooij et al.\,2011) and that these clusters could be the younger counterparts of the old Galactic GCs with multiple populations (Keller et al.\,2011; Conroy \& Spergel\,2011). The possibility that intermediate-age GCs host multiple stellar  populations with different ages has been further supported by the presence of either extended or dual red clumps that indicate that these stellar systems are not consistent with a single isochrone (Girardi et al.\,2009).
An alternative scenario has been proposed by Bastian \& De Mink (2009) who have suggested that the eMSTO is due to stellar rotations in clusters where a single stellar generation exists (see Goudfrooij et al.\,2014, Niederhofer et al.\,2015 and references therein for discussions). Yang et al.\,(2011) have suggested merged binary systems and interacting binaries with mass transfer can be responsible for both the eMSTO and the dual clumps. 

 More recently, an eMSTO has been detected for the first time in the $\sim$300-Myr old cluster NGC\,1856 in the LMC by Milone et al.\,(2015) who have provided the first evidence that this CMD feature is not a peculiarity of intermediate-age clusters (see also Correnti et al.\,2015). Furthermore, NGC\,1856 exhibits a split MS, with the blue MS hosting about one third of the total number of MS stars (Milone et al.\,2015). 

 This finding reveals that young clusters with eMSTO and split MS may provide strong constraints to discriminate between age and rotation. Indeed, rotational models predict that the blue MS would evolve into the faint MSTO,  while in the case of an age spread, in a young cluster we still have a split of the upper MS but the blue MS would be connected to the bright part of the MSTO.
 Unfortunately, the connection between stars along the two MSs and the eMSTO of NGC\,1856 is unclear and the CMD of this cluster has been interpreted in terms of both a difference in age (Milone et al.\,2015; Correnti et al.\,2015) and with two coeval stellar populations with different rotational velocity (D'Antona et al.\,2015).  Noticeably, a split MS has been previously observed in the $\sim$100-Myr old LMC cluster NGC\,1844 (Milone et al.\,2013) but in this case there is no evidence of an eMSTO.

In the present paper we exploit high-precision photometry from the {\it Hubble Space Telescope} of the even younger cluster NGC\,1755 in the LMC.
NGC\,1755, with an age of $\sim$80 Myr, is indeed a particularly interesting object to investigate the morphology of the MS and MSTO in very young clusters.

 The paper is organised as follows. In Sect.~\ref{sec:data} we present the dataset and the methods used to derive stellar photometry and astrometry. The CMD of NGC\,1755 is shown in Sect.~\ref{sec:ms} where we study and characterize the double MS of this cluster. In Sect.~\ref{sec:models} we compare the observed CMD with theoretical models. A summary and discussion is provided in Sect.~\ref{sec:discussion}.

\section{Data and data analysis} \label{sec:data} 
 The main dataset used to investigate the multiple stellar populations in NGC\,1755 consists of images taken with Ultraviolet and Visual Channel of the Wide Field Camera 3 (UVIS/WFC3) on board of {\it HST\,}. Specifically, we have used 2$\times$711s images collected through the F336W band and 90s$+$678s images collected through the F814W band of UVIS/WFC3 as part of GO\,14204 (PI.\,A.\,P.\,Milone).  
 In addition we have used two archive images taken with the Wide Field Channel of the Advanced Camera for Survey (WFC/ACS) of {\it HST} (GO\,9891, PI.\,G.\,Gilmore) through the F555W (50s) and the F814W filters (40s). 

\begin{centering} 
\begin{figure*} 
 \includegraphics[width=8.5cm]{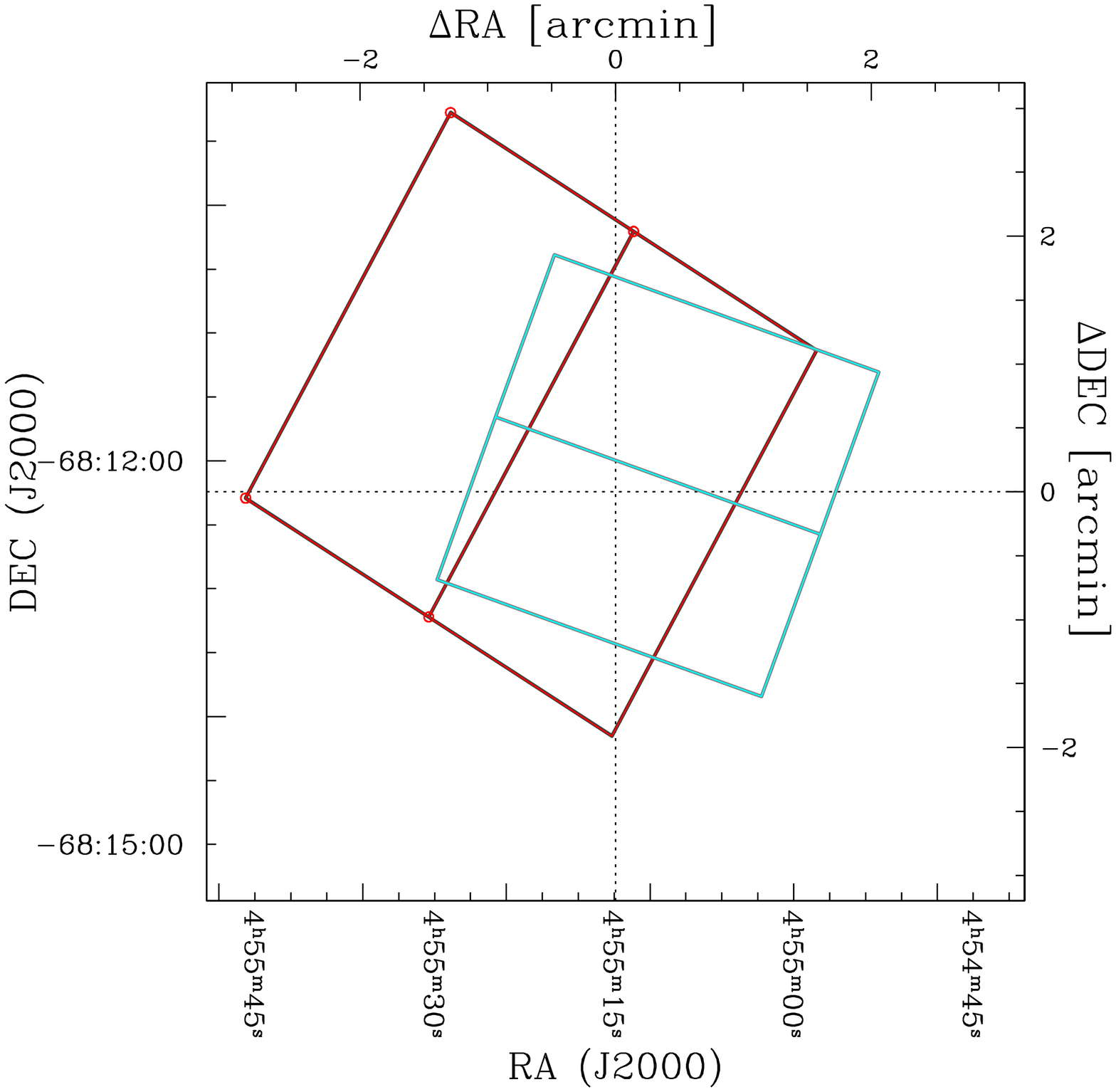} 
 \includegraphics[width=8.5cm]{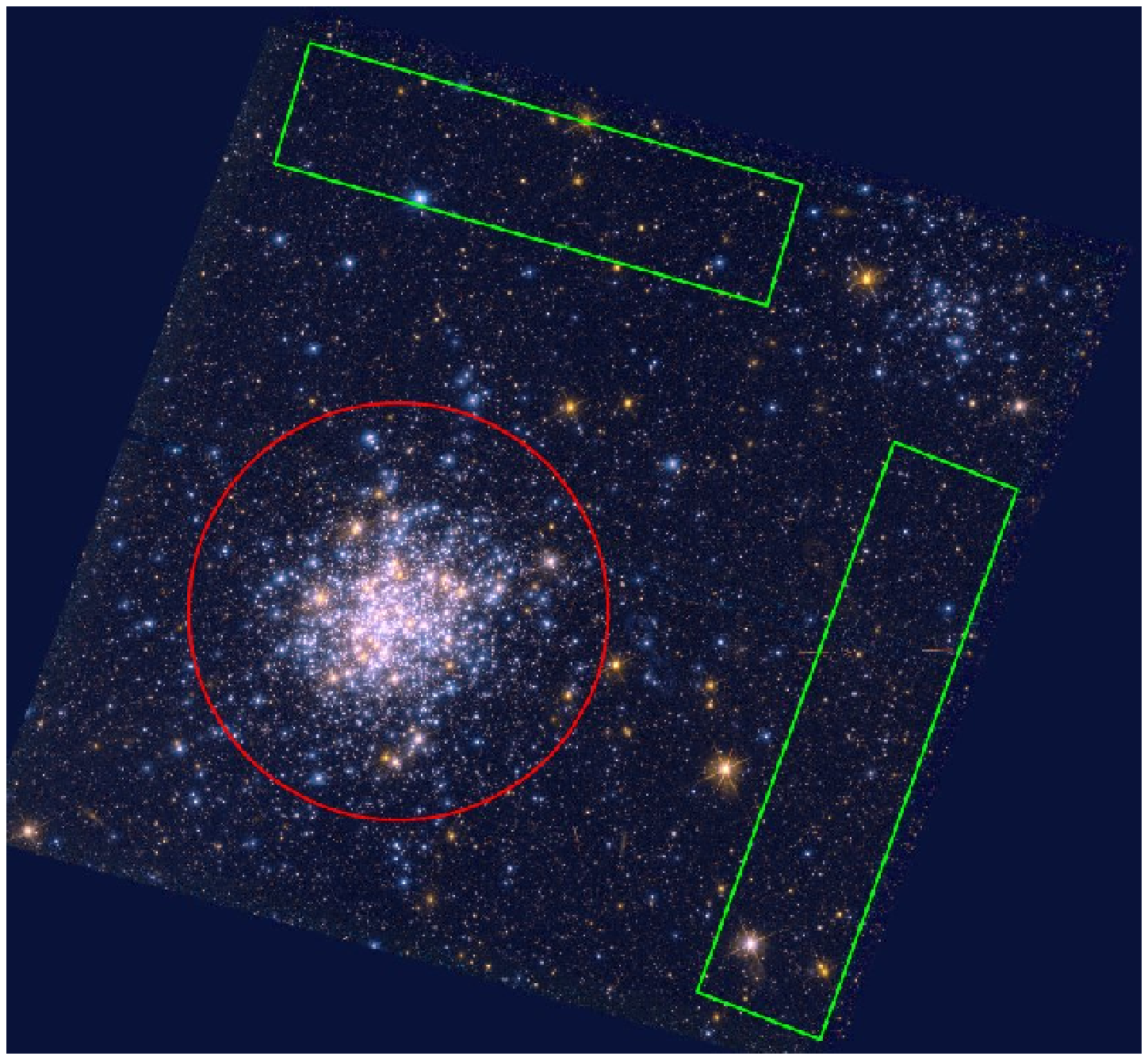} 
%/home/milone/NUBI/WFC3UVIS/NGC1755/footprint/totnero.macro tot 3.2 NGC1755
 \caption{\textit{Left panel:}  Footprints of the data sets used in this work.
  We have used red and the cyan color codes to represent the images taken with WFC/ACS (GO\,9891) and UVIS/WFC3 (GO\,14204), respectively.
 \textit{Right panel:} Stacked trichromatic image of the UVIS/WFC3 field. The red circle marks the NGC\,1755 field, while the two green rectangles mark the reference fields (see text for details). The young stellar cluster NGC\,1749 is visible in the upper-right corner.} 
 \label{fig:footprints} 
\end{figure*} 
\end{centering}

The left panel of Fig.~\ref{fig:footprints} shows the footprints of the WFC/ACS (red) and UVIS/WFC3 (cyan) images used in this paper and includes a field of view of 3.2$\times$3.2 arcmin around NGC\,1755. The right panel of Fig.~\ref{fig:footprints} shows a trichromatic stacked image of the UVIS/WFC3 field. The red circle has a radius of 40 arcsec and delimits a region mostly populated by cluster members that we will call the NGC\,1755 field. The green rectangles mark two regions where the contamination from NGC\,1755 is negligible and have, in total, the same area as the NGC\,1755 field. These two regions will be indicated as reference field and will be used in Sect.~\ref{sec:ms} to statistically estimate the field-star contamination in the NGC\,1755 field. The stellar cluster NGC\,1749, which will be not studied in this paper, is visible in the upper-right corner of the stacked image of the UVIS/WFC3 field.

%CTE
The  ACS/WFC images have been corrected for the poor charge-transfer efficiency (CTE) by using the method and the computer program by Anderson \& Bedin\,(2010).  Similarly, the correction for the poor CTE of UVIS/WFC3 images has been carried out by using a software written by Jay Anderson which is based on the recipe by Anderson \& Bedin\,(2010).

%data reduction ACS+WFC3
We have measured stellar magnitude and positions from the WFC/ACS images by using the computer program from Anderson et al.\,(2008). Specifically, we have measured bright and faint stars by using two different approaches. Bright stars have been measured in each image independently, by fitting to the central 5$\times$5 pixel of each star the appropriate PSF model and the measurements have been combined later. The PSF model has been taken from the library PSF by Anderson \& King\,(2006) plus a constant perturbation that accounts for any focus variation of the spacecraft. The magnitude and position of each very faint star has been determined by fitting all the pixels in all the exposures simultaneously, and by using the same PSF model adopted for bright stars. In order to measure magnitude and flux from the UVIS/WFC3 data we have adapted the program {\it kitchensink} to these images and detector.

For the bright stars with $m_{\rm F336W}<16.55$ or $m_{\rm F814W}<15.50$ we had to derive photometry from their saturated images. To do this we used the method developed by Gilliland (2004), which recovers the electrons that have bled into neighbouring pixels. We refer to the Sect.~8.1 of Anderson et al.\,(2008) for details on our application of this method.
%
%astrometria, calibrazione, pulizia
We have corrected the stellar positions for geometrical distortion by using the solution provided by Bellini, Anderson, \& Bedin (2011) for UVIS/WFC3 and by Anderson \& King (2006) for WFC/ACS. Photometry has been calibrated into the Vega-mag systems as in Bedin et al.\,(2005) and by adopting the zero points provided by the STScI web page for WFC3/UVIS\footnote{http://www.stsci.edu/hst/acs/analysis/zeropoints/zpt.py}. 
We have used several indexes provided by the software as diagnostics of the photometric and astrometric quality (Anderson et al.\,2008) and we have selected a sample of relatively isolated stars with small astrometric and photometric errors, and well fitted by the PSF. We refer to Sect.\,2.1 of Milone et al.\,(2009) for details.

%%%%%%%%%%%%%%%%%%%%%%%%%%%%%%%%%%%%%%%%%%%%%%%%%%%%%%%%%%%%%%%%%%%%%%%%%%
%%%%%%%%%%%%%%%%%%%%%%%%%%%%%%%%%%%%%%%%%%%%%%%%%%%%%%%%%%%%%%%%%%%%%%%%%%

\subsection{Artificial stars}
Artificial stars (ASs) have been used both to estimate the completeness level of each star and to generate synthetic CMDs to be compared with the observations.
 The method adopted to perform the AS-test has been described by Anderson et al.\,(2008). Briefly, we have generated a list of 5$\times$10$^{5}$ ASs, including their coordinates and magnitudes and placed them along the MS of NGC\,1755. For each AS in the list the software developed by Anderson et al.\,(2008) generates a star in the images with the appropriate flux and position and reduces it by using exactly the same procedure and PSF adopted for real stars.
 Moreover, the software provides for ASs the same diagnostics of the photometric and astrometric quality as for real stars. Therefore, we have used in our study of NGC\,1755 only those ASs that are relatively isolated, have small astrometric and photometric errors, and are well fitted by the PSF. 

Completeness has been calculated for each observed star according to its luminosity and radial distance from the cluster center. To do this we have followed the recipe by Milone et al.\,(2009). In summary, we have defined a grid of $9 \times 6$ points in the magnitude versus radius plane and for each point we have determined the completeness between the number of recovered to added ASs. The completeness value associated to each star has been linearly interpolated among these grid points.

\section{The double Main Sequence of NGC\,1755}\label{sec:ms}
%%%%%%%%%%%%%%%%%%%%%%%%%%%%%%%%%%%%%%%%%%%%%%%%%%%%%%%%%%%%%%%%%%%%%%%%%%
\begin{centering} 
\begin{figure*} 
 \includegraphics[width=8.5cm]{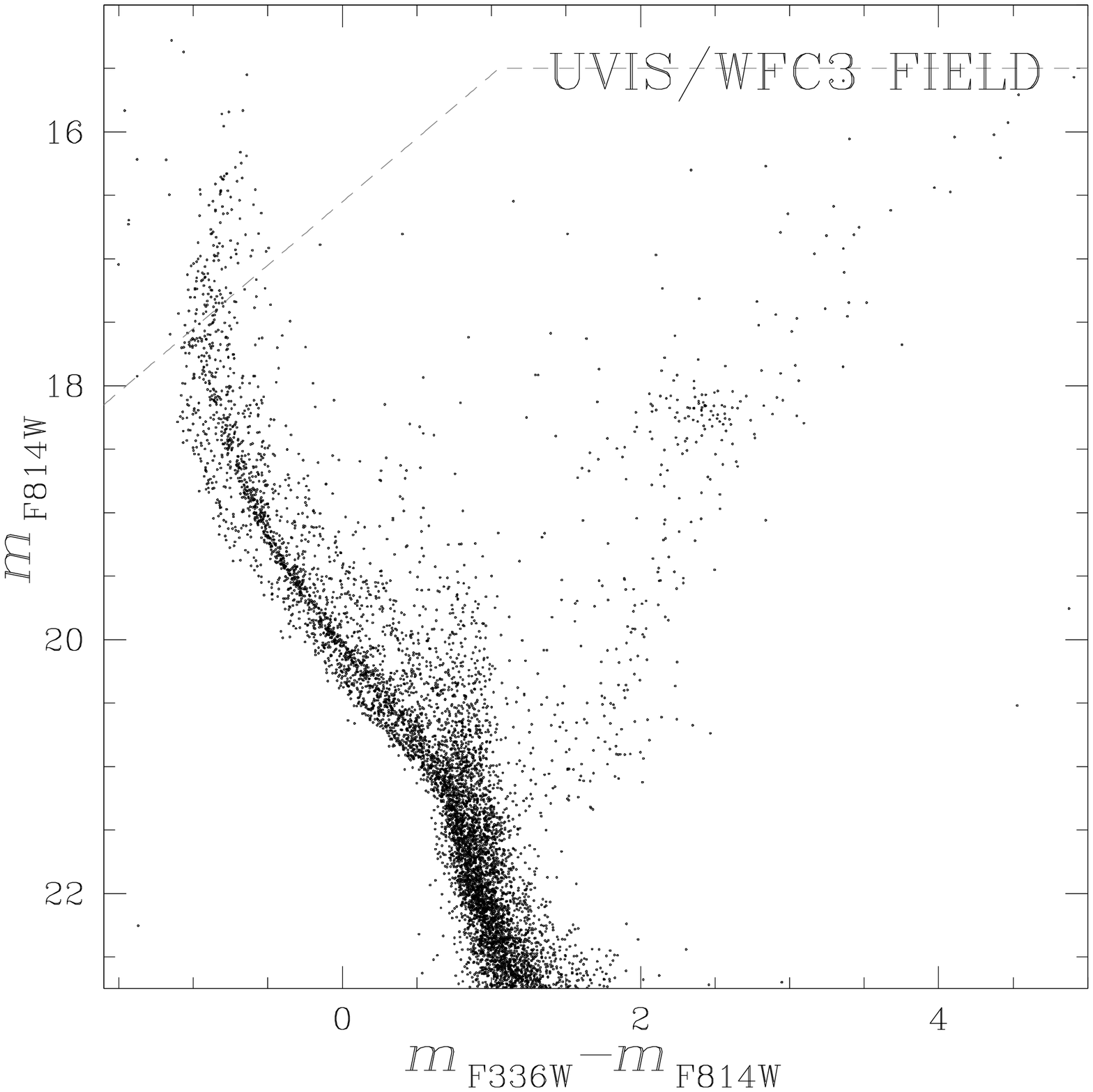} 
 \includegraphics[width=8.5cm]{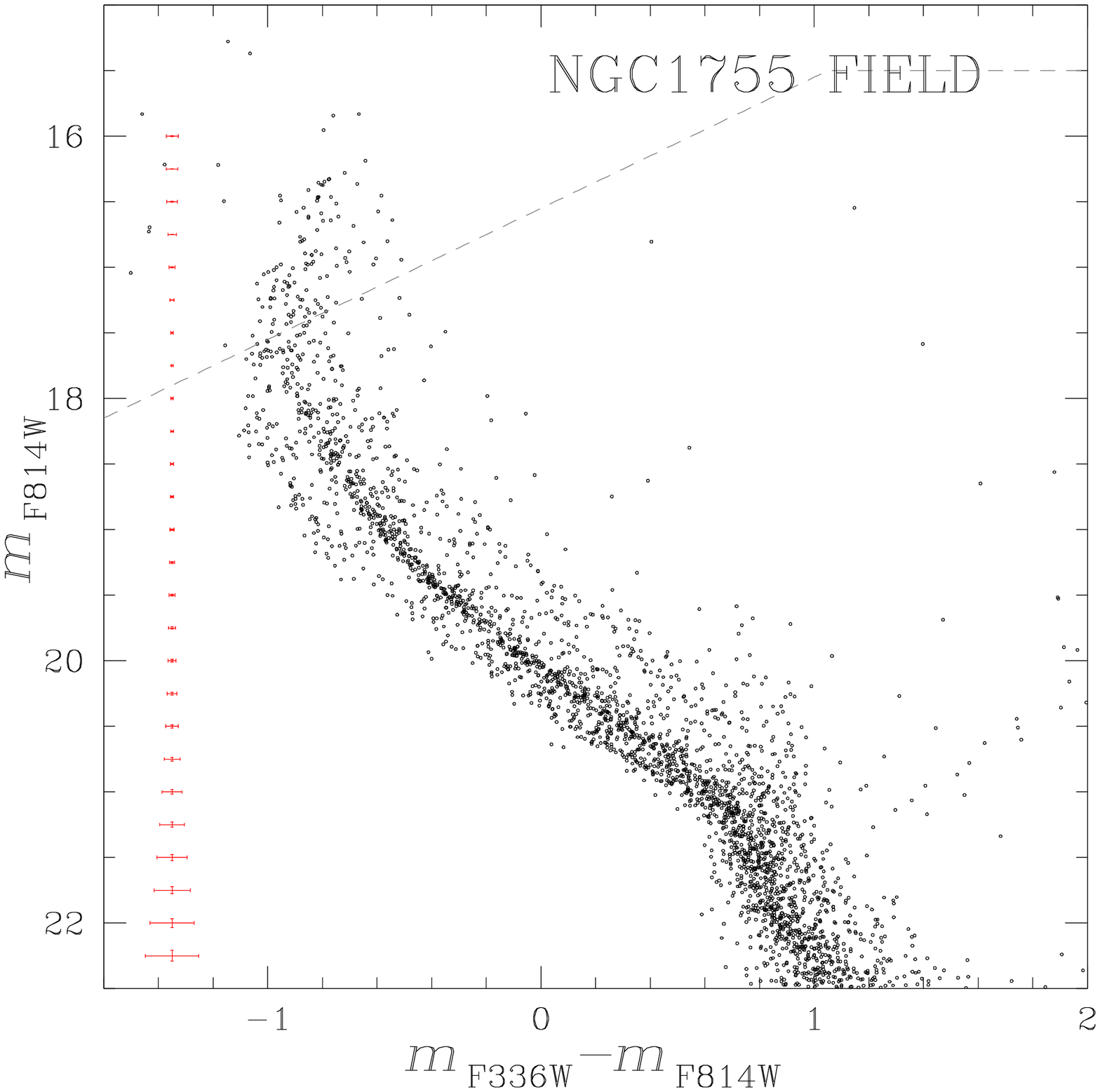} 
%/home/milone/NUBI/WFC3UVIS/NGC1755/figure/fig.macro go0 go (gor regioni) 
 \caption{\textit{Left panel:} $m_{\rm F814W}$ vs.\,$m_{\rm F336W}-m_{\rm F814W}$ CMD for all the stars in the UVIS/WFC3 field of view. \textit{Right panel:} Zoom of the CMD around the MS. Only stars with radial distance smaller than 40 arcsec from the center of NGC\,1755 have been plotted. The photometry of stars above the dashed line has been derived from saturated images.} 
 \label{fig:cmds} 
\end{figure*} 
\end{centering} 
%%%%%%%%%%%%%%%%%%%%%%%%%%%%%%%%%%%%%%%%%%%%%%%%%%%%%%%%%%%%%%%%%%%%%%%%%%

 The left panel of Fig.~\ref{fig:cmds} shows the $m_{\rm F814W}$ vs.\,$m_{\rm F336W}-m_{\rm F814W}$ CMD for all the stars in the UVIS/WFC3 field of view that pass the criteria of selection discussed in Sect.~\ref{sec:data}. The most-prominent feature of the CMD are the two MSs of young stars, that are clearly visible over a range of more than three magnitudes, from $m_{\rm F814W} \sim 17.75$ towards fainter luminosity. Moreover, at colors redder than $m_{\rm F336W}-m_{\rm F814W} \sim 0.5$  the broadened red-giant branch (RGB), sub-giant branch (SGB), and MS of field LMC stars are also visible. 

 The right panel of Fig.~\ref{fig:cmds} is a zoom of the left-panel CMD around the MS, where we have plotted only stars in the NGC\,1755 field of view (red circle in the right panel of Fig.~\ref{fig:footprints}). This figure highlights the main components of the multimodal MS. Specifically, i) a blue and poorly-populated MS, ii) a red MS, containing the majority of MS stars and, iii) a broadened distribution of stars on the red side of the red MS. As we will discuss in  Sect.~\ref{sub:binaries} the latter is made of binary systems formed by pairs of MS stars. A visual inspection of the NGC\,1755 field reveals that the blue MSTO is located around $m_{\rm F814W} \sim 18.0$ while the MSTO of the red MS is approximately 0.5 mag brighter. 
  The red and the blue MS seem to merge together below $m_{\rm F814W} \sim 21.00$ although we can not exclude that a small color difference between the two sequeces, not detectable with the present dataset due to photometric errors, is present at fainter luminosity.

The photometry of the stars brighter than the dashed line in both panels of Fig.~\ref{fig:cmds} has been derived from saturated images.
 The error bars shown in the right panel are indicative of the 1-$\sigma$ uncertainty in color and magnitude. For stars with $m_{\rm F814W}>15.50$ and with $m_{\rm F336W}>16.55$ the errors in the F814W and in the F336W magnitude have been derived from ASs. Specifically, we have divided the CMD into intervals of  0.2 mag, and for each interval we have calculated the difference between the magnitude and the color of the stars in the input list and the recovered magnitudes and colors. The magnitude and color errors have been estimated as the 68.27$^{\rm th}$ percentile of the absolute values of these differences.  The photometric errors of stars with $m_{\rm F814W}<15.50$ or $m_{\rm F336W}<16.55$, whose photometry have been derived from their saturated images, have been estimated as the r.\,m.\,s.\, of the distinct magnitude measurements. 

  In the following we will demonstrate that the three MS components are intrinsic features of the cluster CMD, that can not be attributed neither to field-star contamination, nor to photometric errors and/or differential reddening. Moreover we will show that neither the red, nor the blue MS are sequences made of non-interacting binary stars.  

\subsection{Field-stars contamination}\label{sub:field}
The CMDs in Fig.~\ref{fig:cmds} include both cluster members and field stars.
In order to investigate the CMD of NGC\,1755 members we have statistically subtracted field stars from the CMD of stars in the NGC\,1755 field. 
To do this, we have adopted the procedure used in the previous papers of this series (Milone et al.\,2009, 2013, 2015). 
 Briefly, we have determined for each star (i) in the reference fields a distance  \\ {\scriptsize $d_{\rm i}$=$\sqrt { (k ((m_{\rm F336W, cf}-m_{\rm F814W, cf})-(m_{\rm F336W, rf}^{\rm i}-m_{\rm F814W, rf}^{\rm i})))^{2}   +  (m_{\rm F336W, cf}-m_{\rm F336W, cf}^{\rm i})^{2}}$},\\ where $m_{\rm F336W, cf}$ and $m_{\rm F814W, cf}$ are the magnitudes of stars in the NGC\,1755 field, while the magnitudes of stars in the reference fields are indicated as $m_{\rm F336W, rf}$ and $m_{\rm F814W, rf}$. 
 The constant, $k=3.9$ has been derived as described in detail in Sect.~3.1 of Marino et al.\,(2014) and accounts for the fact that the color of a star is better constrained than its magnitude (Gallart et al.\,2003). 
 Then, we have flagged the closest star in the NGC\,1755 field as a candidate to be subtracted and associated to that star a random number $0<r_{\rm i}<1$. Finally, we have subtracted from the CMD of stars in the NGC\,1755 field all the candidates with $r_{\rm i}< c_{rf}^{\rm i}/c_{cf}^{\rm i}$ where $c_{rf}^{\rm i}$ and $c_{cf}^{\rm i}$ are the values of the completeness of the star (i) in the reference field and the completeness of the closest star in the NGC\,1755 field, respectively.

\begin{centering} 
\begin{figure*} 
 \includegraphics[width=15.5cm]{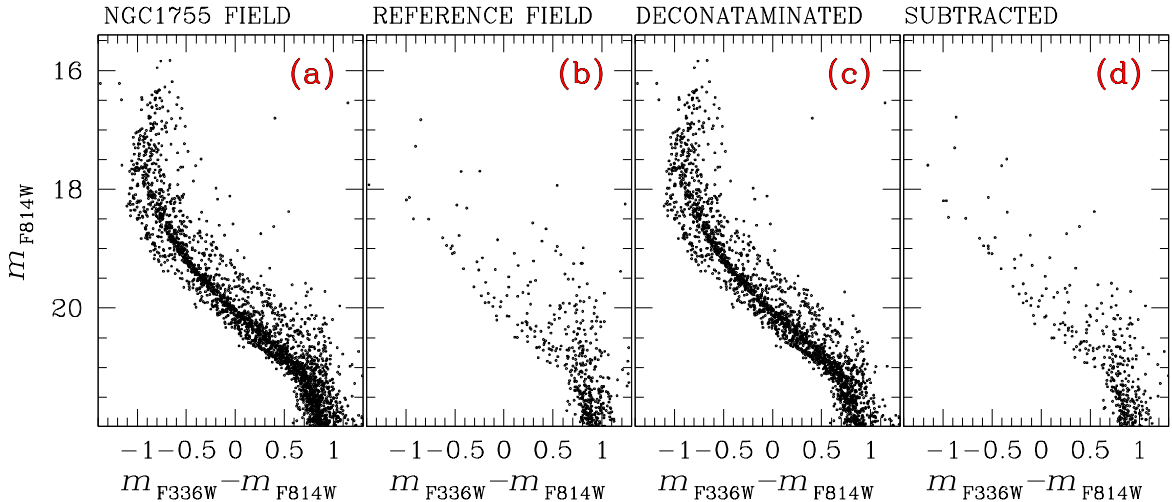} 
%/home/milone/NUBI/WFC3UVIS/NGC1755/figure/fig.macro go2  (gor regioni go0 go) 
 \caption{$m_{\rm F814W}$ vs.\,$m_{\rm F336W}-m_{\rm F814W}$  CMD of stars in the NGC\,1755 field and of stars in the reference field (panels a and b). CMD of stars in the NGC\,1755 field after that the stars in the reference field have been statistically subtracted (panel c). CMD of the subtracted stars (panel d).} 
 \label{fig:decon} 
\end{figure*} 
\end{centering}

The results are illustrated in Fig.~\ref{fig:decon}. Specifically, panels (a) and (b) show the CMD of stars in the NGC\,1755 field and in the reference fields, respectively. The decontaminated CMD is plotted in the panel (c), while the CMD of all the stars that have been statistically subtracted in shown in panel (d).
The fact that the blue and red MS are clearly visible in the decontaminated CMD of panel (c) demonstrates that both sequences belong to NGC\,1755.

%%%%%%%%%%%%%%%%%%%%%%%%%%%%%%%%%%%%%%%%%%%%%%%%%%%%%%%%%%%%%%%%%%%%%%%%%%
%%%%%%%%%%%%%%%%%%%%%%%%%%%%%%%%%%%%%%%%%%%%%%%%%%%%%%%%%%%%%%%%%%%%%%%%%%
\subsection{Spatial variations of the photometric zero point}\label{sub:zp}
% F336W        F814W      F555W        F814W         B           V 
% UVIS        UVIS         ACS          ACS           
%0.51602106  0.205615012  0.339590301  0.203931619    0.426988838 0.326988873
 Variations of the photometric zero point along the field of view can either be due  to differential reddening and to small inaccuracies in the adopted PSF model (Anderson et al.\,2008). Their main effect on the CMD is a broadening of the sequences, and any photometric study of multiple stellar populations would require an appropriate  quantitative assessment of such zero-point variations.

To do this we have first applied to the stars in the NGC\,1755 field the method by Milone et al.\,(2012a) for the correction of differential reddening. As a first step we have determined the absorption coefficient in the UVIS/WFC3 F336W and F814W bands and in the WFC/ACS F555W and F814W bands appropriate for the analysed bright-MS stars of NGC\,1755. 

We have employed the ATLAS12 (Kurucz\,2005; Sbordone et al.\,2007) and SYNTHE codes (Kurucz\,2005) to perform the spectral synthesis in the wavelength interval between 2,500 and 10,000 \AA\, of a star with effective temperature, $T_{\rm eff}$=13,285 K, gravity log${g}=4.29$, and metallicity Z=0.006. The resulting spectrum will be indicated as reference. According to isochrones from the Geneva database\footnote{http://obswww.unige.ch/Recherche/evoldb/index/} (Georgy et al.\,2014), the adopted stellar parameters correspond to a MS star of NGC\,1755 with $m_{\rm F814W}=19.1$ once we assume reddening $E(B-V)$=0.1 and distance modulus (m$-$M)$_{0}$=18.4 (see Sect.~\ref{sec:models} for details). 
Then, we have  determined an absorbed spectrum by convolving the flux of reference spectrum with the extinction law by Cardelli et al.\,(1989). In order to do this we have assumed E(B$-$V)=0.1 and $R_{\rm V}=3.1$. 

\begin{centering} 
\begin{figure*} 
 \includegraphics[width=15.5cm]{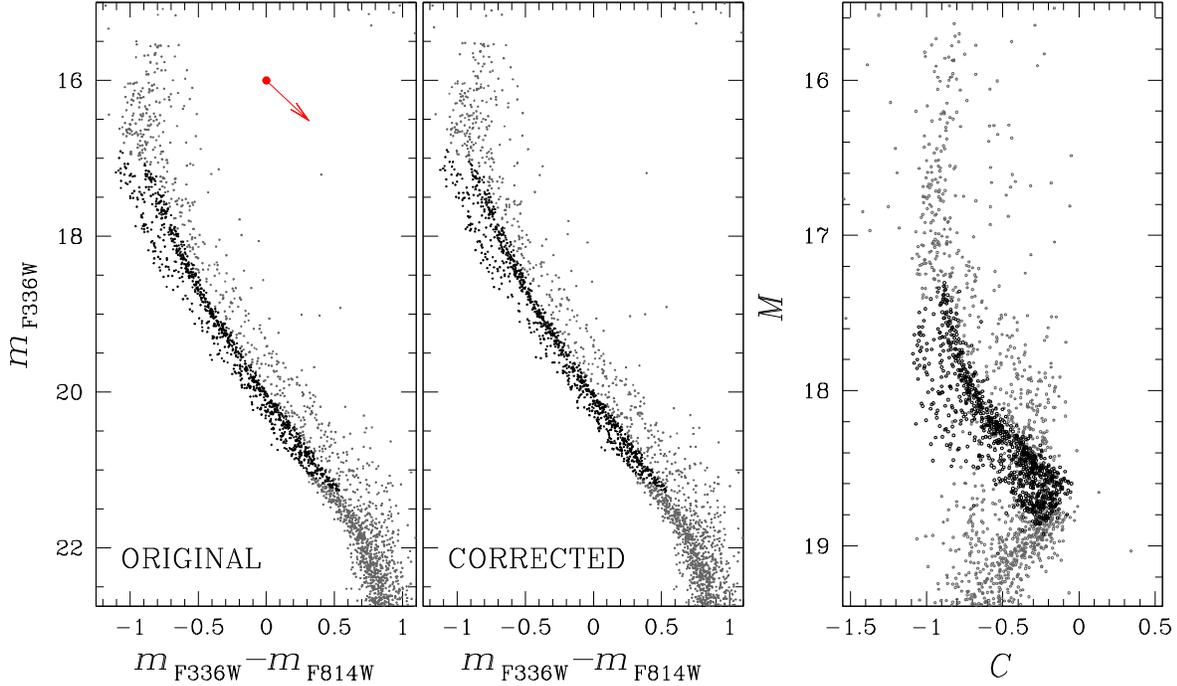} 
%/home/milone/NUBI/WFC3UVIS/NGC1755/figure/fig.macro go6
%/home/milone/NUBI/WFC3UVIS/NGC1755/SYNTHSPECTRA/leggered.macro  
 \caption{Original $m_{\rm F336W}$ vs.\,$m_{\rm F336W}-m_{\rm F814W}$  CMD of stars in the NGC\,1755 field (left panel), and CMD corrected for differential reddening (middle panel). The red arrow in the left-panel indicates the direction of the reddening vector and its length corresponds to a reddening variation of $\Delta$E(B$-$V)=0.1 mag. The right panel shows the reddening-free pseudo-CMD $M$ vs.\,$C$ CMD, where $C=$($m_{\rm F336W}-m_{\rm F555W}$)$-$1.316 ($m_{\rm F555W}-m_{\rm F814W}$) and $M= m_{\rm F814W}+0.914\,m_{\rm F555W}-m_{\rm F336W}$. Black dots represent MS stars with $17.95<m_{\rm F814W}<20.75$ where the split MS is more evident.} 
 \label{fig:reddening} 
\end{figure*} 
\end{centering}

The two synthetic spectra have been finally integrated over the transmission curves of the filters used in this paper to derive the corresponding magnitudes. From the comparison of the magnitude of the absorbed and the reference spectrum  we find the following values for the absorption coefficients: $A_{\rm F336W}^{\rm UVIS/WFC3}=5.16$ E(B$-$V), $A_{\rm F814W}^{\rm UVIS/WFC3}=2.06$ E(B$-$V), $A_{\rm F555W}^{\rm WFC/ACS}=3.40$ E(B$-$V), and $A_{\rm F814W}^{\rm WFC/ACS}=2.04$ E(B$-$V).
 These coefficients have been used to derive the direction of the reddening vector that we have indicated as a red arrow in the $m_{\rm F336W}$ vs.\,$m_{\rm F336W}-m_{\rm F814W}$ CMD of stars in the NGC\,1755 field shown in the left panel of Fig.~\ref{fig:reddening}.

In order to determine the differential reddening in the NGC\,1755 field, we have first rotated the CMD in a reference frame where the abscissa is parallel to the reddening direction. Then we have determined the fiducial line along the red MS and selected a sample of reference red-MS stars with high-precision photometry that pass the criteria of selection of Sect.~\ref{sec:data}. For each star in the NGC\,1755 field we have selected a sample of 50 neighbour reference stars and calculated the residual colors with respect to the fiducial line. The median of these residuals, determined along the reddening direction, has been assumed as the  differential reddening corresponding to each star. The corresponding error has been estimated as the r.m.s of the 50 residual colors divided by the square root of 49. We have carefully excluded any star from the determination of its own differential reddening. 

 The variation of reddening in the NGC\,1755 field of view is very small, the $68.27^{\rm th}$ percentile of the differential-reddening distribution is $\Delta$E(B$-$V)=0.003 mag and is comparable with the corresponding uncertainty of 0.002 mag. The reddening variation never exceeds $\Delta$E(B$-$V)=$\pm$0.008 mag. 
 Results are illustrated in Fig.~\ref{fig:reddening} where we compare the original (left panel) and  the corrected CMD (right panel). The two CMDs are very similar and the split MS is clearly visible in both of them thus demonstrating that it is real. Since the reddening variation is negligible, for simplicity, in the following we will use the original photometry.

To further demonstrate that the split MS is not due to differential reddening, we have defined the reddening-free pseudo color $C=$($m_{\rm F336W}-m_{\rm F555W}$)$-$1.316 ($m_{\rm F555W}-m_{\rm F814W}$) and the reddening-free pseudo-magnitude $M= m_{\rm F814W}+0.914\,m_{\rm F555W}-m_{\rm F336W}$. The fact that the two MSs are clearly visible in the $M$ vs.\,$C$ pseudo-CMD plotted in the right-panel CMD further demonstrates that the MS split is real. Moreover, we note that the blue MS is more broadened in the pseudo color $C$ than the red MS.
%%%%%%%%%%%%%%%%%%%%%%%%%%%%%%%%%%%%%%%%%%%%%%%%%%%%%%%%%%%%%%%%%%%%%%%%%%
%%%%%%%%%%%%%%%%%%%%%%%%%%%%%%%%%%%%%%%%%%%%%%%%%%%%%%%%%%%%%%%%%%%%%%%%%%

\subsection{Population ratio}\label{sub:pratio}
\begin{centering} 
\begin{figure*} 
 \includegraphics[width=13.5cm]{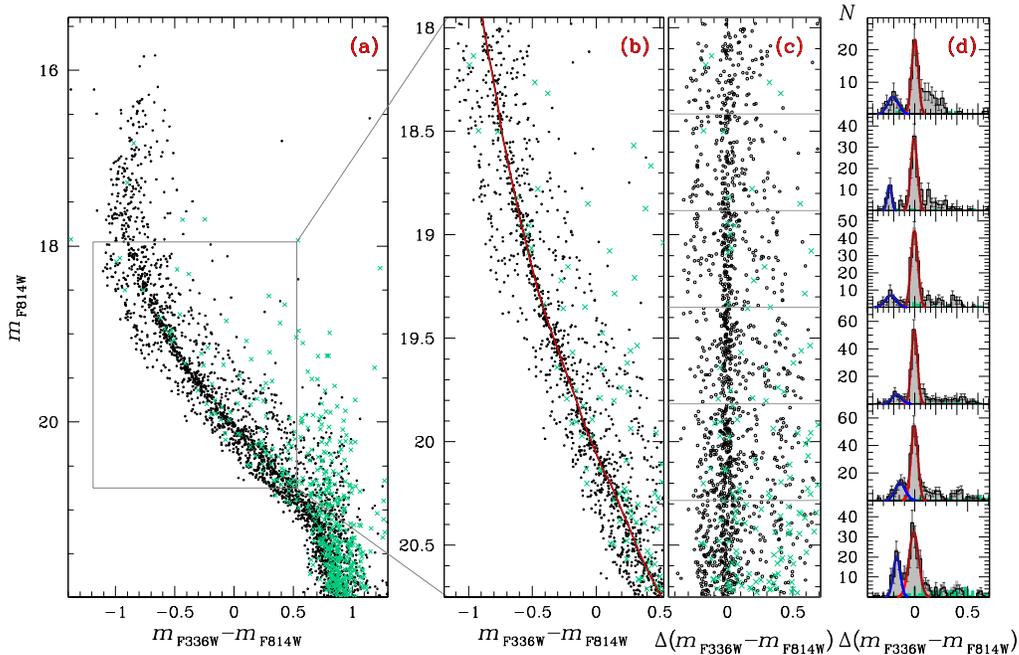} 
%/home/milone/NUBI/WFC3UVIS/NGC1755/figure/fig.macro go4
 \caption{This figure illustrates the procedure to determine the fraction of stars in the blue and the red MS. Panel (a) shows the CMD for both stars in the NGC\,1755 field (black circles) and in the reference fields (aqua crosses). Panel (b) is a zoom of the previous CMD around the region where the double MS is more clearly visible. The red line is the fiducial line of the red MS. The verticalized CMD is plotted in panel (c) for the stars shown in panel (b). Panels (d) show the histogram distribution of all the panel-c stars in six magnitude bins. Specifically the grey- and aqua-shaded histograms refer to stars in the NGC\,1755 and in the reference field, while the black histogram is obtained as the difference of the grey- and the aqua-shaded histograms. The black curve overimposed on the black histogram is the best-fitting function made by the sum of the Gaussians, whose components are coloured red and blue, respectively.} 
 \label{fig:pratio} 
\end{figure*} 
\end{centering}

In order to determine the fraction of stars in each MS we have adopted the method already used in our previous papers and illustrated in Fig.~\ref{fig:pratio} for NGC\,1755. The aqua crosses and the black circles in the CMDs of panels (a) and (b) in Fig.~\ref{fig:pratio} represent stars in the NGC\,1755 and in the reference fields, respectively. Panel b shows a zoom of the CMD for MS stars with $17.95<m_{\rm F814W}<20.75$. The red fiducial line superimposed on the red MS has been determined as follows. We have first selected by eye a sample of bona-fide red-MS stars and calculated their median colors and magnitude in a series of magnitude intervals determined by using the naive-estimator method (Silverman 1986). Specifically, we started to divide the  MS into a series of F814W magnitude bins of width $\nu=0.2$ mag. The bins are defined over a grid of $N$ points separated by steps of fixed magnitude ($s=\nu/3$).
The values of the median colors and magnitude have been smoothed by using the technique of boxcar averaging, where each point has been replaced by the average of the three adjacent points.   

The verticalized $m_{\rm F814W}$ vs.\,$\Delta$($m_{\rm F336W}-m_{\rm F814W}$) diagram shown in panel (c) has been obtained by subtracting from the color of each star, the corresponding color of the fiducial at the same F814W magnitude. We have then divided the analysed interval of magnitude into six bins and determined for each bin the completeness-corrected $\Delta$($m_{\rm F336W}-m_{\rm F814W}$) histogram distribution of stars in the NGC\,1755 field (gray-dashed histogram in panels d) and for stars in the reference fields (aqua-dashed histogram in panels d). Finally, we have determined the black histograms shown in panel (d) as the difference between the gray and the aqua ones and fitted them with a bi-Gaussian function. In order to minimize the effect of binaries, we have used only the histogram region with $\Delta$($m_{\rm F336W}-m_{\rm F814W}$)$<0.1$ in the determination of the best-fit function. The number of blue-MS and red-MS stars has been determined from the area under the red and the blue Gaussians shown in panels (d) of Fig.~\ref{fig:pratio} which are the two components of the best-fitting bi-Gaussian function. 

From the average of the results from the six magnitude bins, we found that 74$\pm$3\% and 26$\pm$3\% of stars belong to the red- and to the blue-MS, respectively. The error has been estimated as the r.\,m.\,s.\,of the six independent estimates of the population ratio ($\sigma=6$\%) divided by the square root of five.
%%%%%%%%%%%%%%%%%%%%%%%%%%%%%%%%%%%%%%%%%%%%%%%%%%%%%%%%%%%%%%%%%%%%%%%%%%
%%%%%%%%%%%%%%%%%%%%%%%%%%%%%%%%%%%%%%%%%%%%%%%%%%%%%%%%%%%%%%%%%%%%%%%%%%

\subsection{Binaries}\label{sub:binaries}

\begin{centering} 
\begin{figure*} 
 \includegraphics[width=10.5cm]{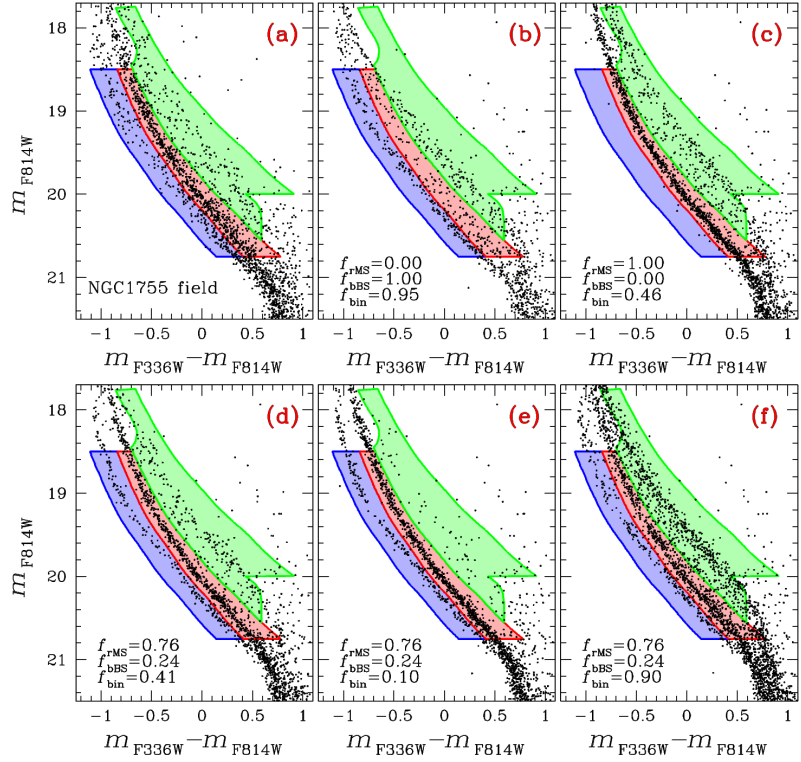} 
%/home/milone/NUBI/WFC3UVIS/NGC1755/figure/fig.macro go8  (go7)
 \caption{Comparison between the observed CMD of stars in the NGC\,1755 field (panel a) and five simulated CMDs corresponding to different fraction of red-MS and blue-MS stars and different binary fraction (panels b--e). The blue, red, and green shaded regions overimposed on each CMD correspond to regions A, B, and C, respectively (see text for details).} 
 \label{fig:binarie} 
\end{figure*} 
\end{centering}

In order to investigate the population of binary stars in NGC\,1755 we have extended to this cluster the method introduced by Milone et al.\,(2012b) in their study of NGC\,2808.
Briefly, we have divided the CMD of stars in the NGC\,1755 field into three regions, namely A, B, and C that we have coloured blue, red, and green, respectively in Fig.~\ref{fig:binarie}a and we have determined the number of stars, corrected for completeness, in each region ($N_{\rm A, B, C}^{\rm obs}$). Regions A and B include most of the single stars in the blue and red MS, while region C is mostly populated by binaries with large mass ratios. 
We have used ASs to generate a large number of CMD with different values for the fraction of blue-MS stars, red-MS stars, and binaries ($f_{\rm bMS}$, $f_{\rm rMS}$, and $f_{\rm bin}$) and compared the simulated and the observed CMDs. We have assumed a flat mass-ratio distribution for binaries as observed in the Galactic GCs for $q>0.5$ (Milone et al.\,2012a; 2016), and, in order to account for field-star contamination in the NGC\,1755 field, the stars in the reference field have been added to each simulated CMD.

We started to investigate the possibility that the split MS is due to non-interacting binaries, whose two components belong both to the blue MS. To do this we have simulated a grid of CMDs with $f_{\rm rMS}$, and $f_{\rm bin}$ ranging from 0.00 to 1.00 in steps of 0.01. The number of stars in the simulated CMD has been normalised in such a way that the region A of the simulated CMD hosts the same number of stars as the observed ones. We found that the observed and simulated CMDs have different morphologies and that the numbers of simulated and observed stars in both the regions B and C are very different regardless the adopted value of $f_{\rm bin}$. Results are shown in the panel (b) of Fig.~\ref{fig:binarie} for the case of $f_{\rm bin}=0.95$. 
Similarly, we have investigated the possibility that the split MS is due to non-interacting red-MS-red-MS binaries. In this case, as shown in Fig.~\ref{fig:binarie}c, the observed and the simulated CMDs have the same number of stars in the regions B and C for $f_{\rm bin}=0.46$. However, in this scenario only a few stars are expected in the region A of the simulated CMD.
These findings demonstrate that the double MS can not be explained by non-interacting binaries only and that a large fraction of binaries formed by red MS stars are needed to reproduce the observations.

In order to determine the fraction of binaries in NGC\,1755 and to obtain a new estimate of the fraction of blue-MS and red-MS stars, we have simulated a grid of CMDs with $f_{\rm bMS}>0$, $f_{\rm rMS}>0$ and $f_{\rm bin}$ ranging from 0.00 to 1.00 in steps of 0.01. For simplicity, we have assumed that the components of each binary system have the same probability to belong both to the red and the blue MS. Moreover, we have imposed that each simulated CMD hosts the same number of stars in both regions A and B as the observed ones.
 We find that the number of stars in the regions A, B, and C of simulated and the observed CMD are the same for $f_{\rm bMS}=0.24 \pm 0.03$, $f_{\rm rMS}=0.76 \pm 0.03$, $f_{\rm bin}=0.41 \pm 0.03$. The derived fraction of blue-MS and red-MS stars are in agreement with those derived in the Sect.~\ref{sub:pratio}. 
 The derived value of $f_{\rm bin}$ demonstrated that NGC\,1755 hosts a large fraction of binaries in close analogy with what is observed in other young and intermediate-age stars MC clusters (e.g.\,Milone et al.\,2009, 2013, 2015; Geller et al.\,2015).

%%%%%%%%%%%%%%%%%%%%%%%%%%%%%%%%%%%%%%%%%%%%%%%%%%%%%%%%%%%%%%%%%%%%%%%%%%
%%%%%%%%%%%%%%%%%%%%%%%%%%%%%%%%%%%%%%%%%%%%%%%%%%%%%%%%%%%%%%%%%%%%%%%%%%

\subsection{The CMD from WFC/ACS photometry}\label{sub:ACS}
\begin{centering} 
\begin{figure*} 
 \includegraphics[width=15.5cm]{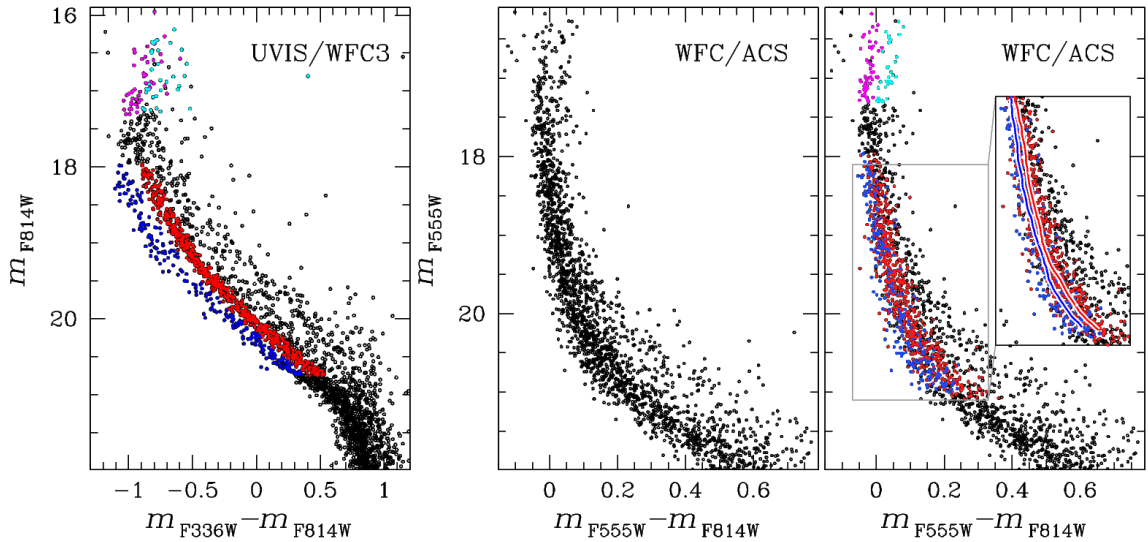} 
%/home/milone/NUBI/WFC3UVIS/NGC1755/figure/fig.macro go5
 \caption{Reproduction of the  CMD of stars in the NGC\,1755 field plotted in the left panel of Fig.~\ref{fig:cmds} and obtained from UVIS/WFC3 photometry (left panel). Middle and right panels show the $m_{\rm F555W}$ vs.\,$m_{\rm F555W}-m_{\rm F814W}$  CMD for stars in the NGC\,1755 field from WFC/ACS photometry. The blue-MS and red-MS stars, selected from UVIS/WFC3 photometry, are coloured red and blue, respectively, in both the left- and the right-panel CMD. Cyan and magenta colors indicate the two group of MSTO-A, and MSTO-B stars selected in the right-panel CMD. The red and the blue lines superimosed on the CMD in the inset are the fiducial lines of the two MSs.} 
 \label{fig:ACS} 
\end{figure*} 
\end{centering}

To further investigate the double MS of NGC\,1755 we compare in Fig.~\ref{fig:ACS} the CMD obtained from UVIS/WFC3 photometry in F336W and F814W bands (left panel) and the $m_{\rm F555W}$ vs.\,$m_{\rm F555W}-m_{\rm F814W}$ CMD from WFC/ACS data (middle and right panel).
 We have used the $m_{\rm F814W}$ vs.\,$\Delta$($m_{\rm F336W}-m_{\rm F814W}$) diagram of Fig.~\ref{fig:pratio}c to select a group of bona-fide red-MS stars with $-0.10\leq\Delta$($m_{\rm F336W}-m_{\rm F814W}<0.07$) and a group of bona-fine blue-MS stars with $-0.40\leq\Delta$($m_{\rm F336W}-m_{\rm F814W}$)$<-0.10$. These red- and blue-MS stars are coloured red and blue, respectively in the left and the right panel of Fig.~\ref{fig:ACS}. We have determined the fiducial lines of the two MSs by using the same method described in the Sect.~\ref{sub:pratio} and show them in the inset of the right-panel of Fig.~\ref{fig:ACS}. We find that blue-MS stars have, on average, bluer $m_{\rm F555W}-m_{\rm F814W}$ color than red-MS stars.

The $m_{\rm F555W}$ vs.\,$m_{\rm F555W}-m_{\rm F814W}$ CMD and the $m_{\rm F814W}$ vs.\,$m_{\rm F336W}-m_{\rm F814W}$ shown in Fig.~\ref{fig:ACS} have been derived from two distinct datasets collected through different filters and cameras. 
If the double MS  of NGC\,1755, identified from UVIS/WFC3 photometry, is an artefact due to photometric errors, we would have the same probability for red-MS stars of being either bluer or redder than the blue in the ACS/WFC CMD.
 The fact that blue-MS and red-MS stars have different colors in both CMD further demonstrates that the double MS is intrinsic (Anderson et al.\,2009).
  Nevertheless, the separation between the blue and the red MS is less evident in $m_{\rm F555W}-m_{\rm F814W}$ than in $m_{\rm F336W}-m_{\rm F814W}$. This fact suggests that at a fixed magnitude, blue-MS and red-MS stars have different effective temperature. Indeed for a given temperature difference, a larger color baseline like $m_{\rm F336W}-m_{\rm F814W}$ corresponds to a wide color difference, while a smaller color baseline, like $m_{\rm F555W}-m_{\rm F814W}$, is responsible for a smaller color change.

A visual inspection at the $m_{\rm F555W}$ vs.\,$m_{\rm F555W}-m_{\rm F814W}$ CMD reveals that, above the MSTO, the MS of NGC\,1755 is broadened with some hints of bimodality. The two MSTO components are separated by $\sim$0.05 mag in the F555W-F814W color and the observed broadening is much larger than that expected from photometric errors only, that for these bright MS stars is smaller than 0.02 mag in color. Similarly, as discussed in Sect.~\ref{sub:zp}, the spatial variations of reddening across the field of view are very small and can not account for the bifurcated MS of NGC\,1755. 
 In the right panel of Fig.~\ref{fig:ACS} we have used cyan and magenta colors to represent the two groups of MSTO-A and MSTO-B stars that we have selected by eye in this CMD. We have used the same color codes to represent these stars in the left-panel CMD.
 We find that, on average MSTO-A stars have redder $m_{\rm F336W}-m_{\rm F814W}$ colors than MSTO-B stars and that both MSTO-A and MSTO-B stars have broadened color distribution in the $m_{\rm F814W}$ vs.\,$m_{\rm F336W}-m_{\rm F814W}$ CMD.
%%%%%%%%%%%%%%%%%%%%%%%%%%%%%%%%%%%%%%%%%%%%%%%%%%%%%%%%%%%%%%%%%%%%%%%%%%
%%%%%%%%%%%%%%%%%%%%%%%%%%%%%%%%%%%%%%%%%%%%%%%%%%%%%%%%%%%%%%%%%%%%%%%%%%

\section{Comparison with theory}\label{sec:models}
In order to understand the physical reasons that are responsible for the double MS and the eMSTO of NGC\,1755 we have compared the observed CMD with isochrones.  We have first investigated the possibility that the observed CMD is consistent with multiple stellar populations with the same metallicity but different ages.
 To do this, we have used isochrones for non-rotating stars from the BaSTI database\footnote{http://www.oa-teramo.inaf.it/BASTI} (Pietrinferni et al.\,2006) which account for the core-convective overshoot during the central H-burning phase (see Pietrinferni et al.\,2004 for details). 

In the scenario where the split MS is due to age variation, the blue MS would correspond to the  second stellar generation, while  first-generation stars would belong to the red MS. We started to determine the best fit between the isochrones and the blue MS.

%%%%%%%%%%%%%%%%%%%%%%%%%%%%%%%%%%%%%%%%%%%%%%%%%%%%%%%%%%%%%%%%%%%%%%%%%%
\begin{centering} 
\begin{figure*} 
 \includegraphics[width=13.5cm]{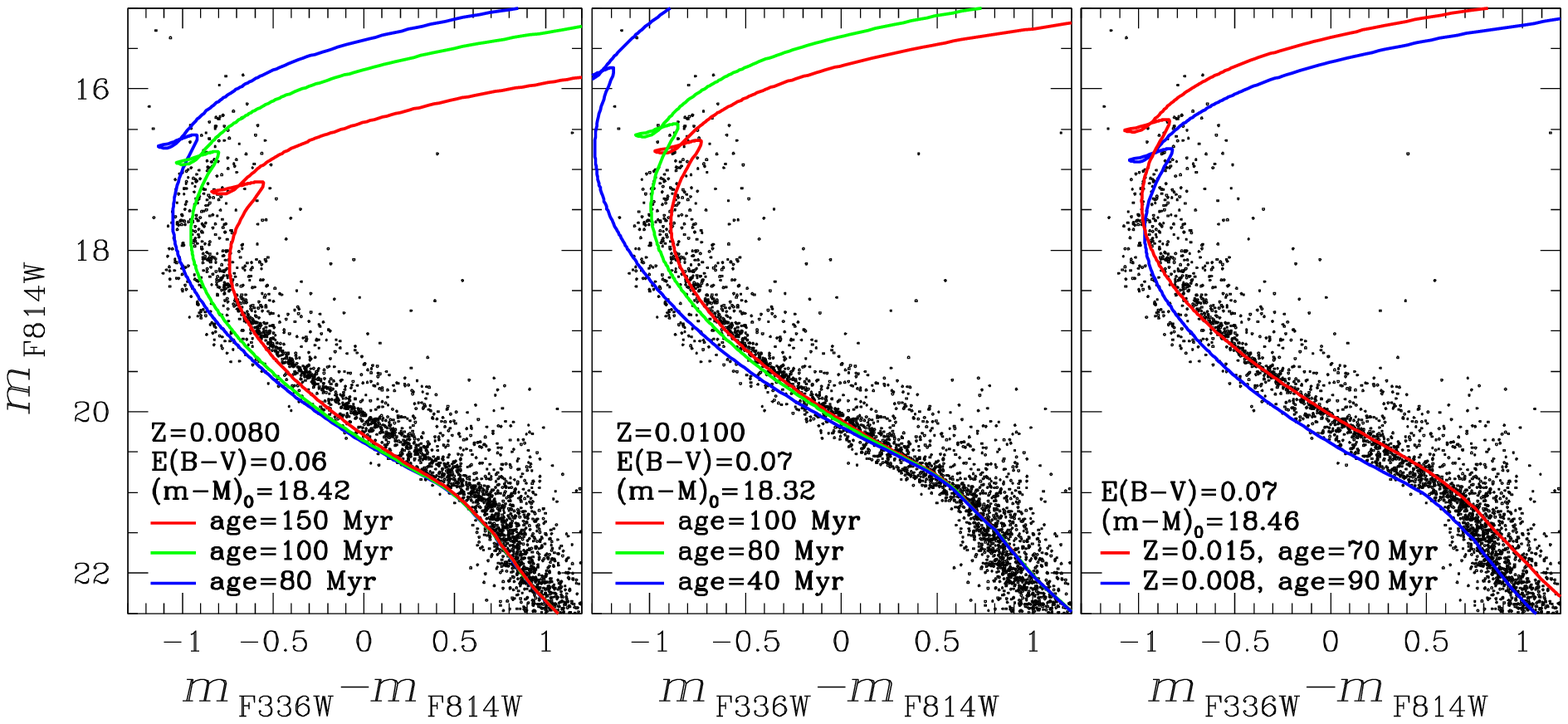} 
%/home/milone/NUBI/WFC3UVIS/NGC1755/figure/fig.macro go9rep1
 \caption{ Left and middle panel show the comparison between the CMD of stars in the NGC\,1755 field and  non-rotating isochrones from the BaSTI database with the same metallicity and helium content but different ages. In the right panel we use non-rotating BaSTI isochrones with the same helium abundance but different metallicity and age. The values adopted for distance modulus, reddening, age, metallicity, and rotation rate are quoted in each panel.} 
 \label{fig:isocrone1} 
\end{figure*} 
\end{centering} 
%%%%%%%%%%%%%%%%%%%%%%%%%%%%%%%%%%%%%%%%%%%%%%%%%%%%%%%%%%%%%%%%%%%%%%%%%%

%%%%%%%%%%%%%%%%%%%%%%%%%%%%%%%%%%%%%%%%%%%%%%%%%%%%%%%%%%%%%%%%%%%%%%%%%%
\begin{centering} 
\begin{figure} 
 \includegraphics[width=9cm]{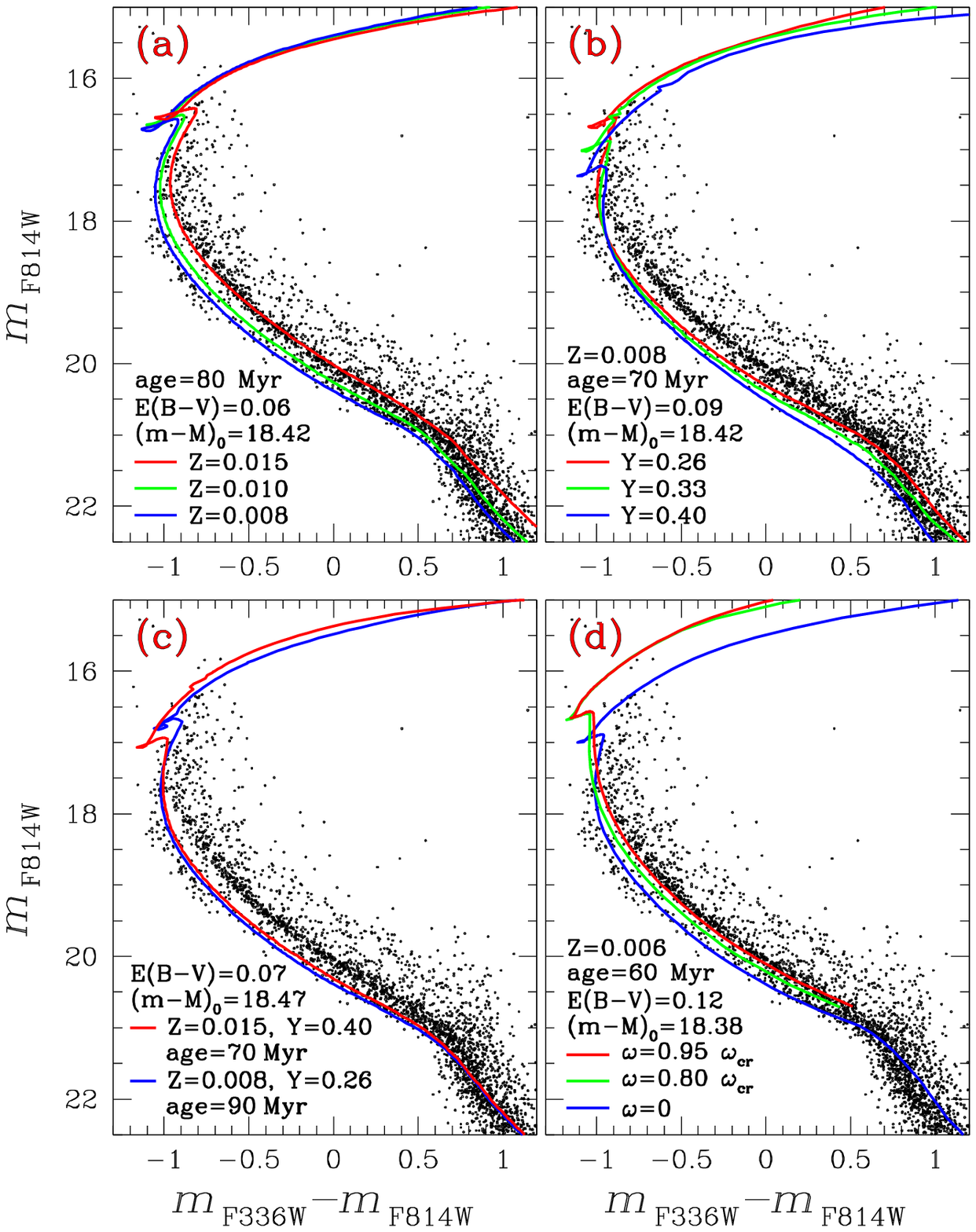} 
%/home/milone/NUBI/WFC3UVIS/NGC1755/figure/fig.macro go9n
 \caption{Comparison between the CMD of stars in the NGC\,1755 field and isochrones from the BaSTI (Pietrinferni et al.\,2006, panel a), Padova (Girardi et al.\,2002, panels c and d) and the Geneva database (Georgy et al.\,2014, panel d). Specifically, in panel (a) we use non-rotating isochrones with the same age but different metallicity and in panel (b) we show isochrones with the same age and metallicity but different helium abundance. Panel (d) shows isochrones with different metallicity, helium and age.
 In panel (d) we plot isochrones with the same age, helium and metallicity but different rotation rates. The values adopted for distance modulus, reddening, age, metallicity, and rotation rate are quoted in each panel.} 
 \label{fig:isocrone2} 
\end{figure} 
\end{centering} 
%%%%%%%%%%%%%%%%%%%%%%%%%%%%%%%%%%%%%%%%%%%%%%%%%%%%%%%%%%%%%%%%%%%%%%%%%%

 The left panel of Fig.~\ref{fig:isocrone1} shows that the BaSTI isochrone with metallicity Z=0.008, helium abundance Y=0.26, and age of 80 Myr (blue line) provides a good match with the blue MS when we assume reddening E(B$-$V)=0.06, distance modulus, (m$-$M)$_{0}$=18.42 and adopt the relation between the reddening and the absorption coefficients that we have derived in Sect.~\ref{sub:zp}. We also show in  the right panel of Fig.~\ref{fig:isocrone1} two isochrones with the same metallicity and helium abundance but ages of 100 Myr (green line) and 150 Myr (red line). From a visual comparison between these isochrones and the observed CMD it is clear that,  while age variation alone  reproduces quite well the eMSTO, it cannot be the responsible for the split MS of NGC\,1755. 
%%%  

 As an alternative approach, in the central panel of Fig.~\ref{fig:isocrone1}, we show that the green and the red isochrones with Z=0.010, Y=0.26, and age of 80 and 100 Myr, respectively, provide a good match with the red MS when we assume E(B$-$V)=0.07 and (m$-$M)$_{0}$=18.32. We find that a younger isochrone with age of 40 Myr, provides a good fit of the blue MS for $m_{\rm F814W} \lesssim 20.5$ but it does not reproduce neither the MSTO of NGC\,1755 nor the fainter part of the blue MS. We conclude that age variation is not responsible for the CMD morphology of NGC\,1755.

In  the right panel of Fig.~\ref{fig:isocrone1} we have compared the observed CMD of NGC\,1755 with two isochrones with different metallicity and age, with the most metal-rich isochrone being also the youngest one. The combination of age and metallicity variations provides a better match of the split MS, when compared to the case of a age variation alone shown in the left and middle panel.  Nevertheless, we still get a poor fit of the faint MS with $21.00 \lesssim m_{\rm F814W}\lesssim 22.25$.

In panel (a) of Fig.~\ref{fig:isocrone2} we investigate the possibility that the observed CMD of NGC\,1755 is consistent with distinct stellar populations with the same age but different metallicity. In this case, the blue MS would correspond to the most metal-poor stellar population.
The blue isochrone superimposed on the panel-(a) CMD has metallicity, Z=0.0080 and is the same isochrone shown in the left panel of Fig.~\ref{fig:isocrone2}. The green and the red isochrones have the same age and helium abundance as the blue one but a metallicity of Z=0.010 and Z=0.015, respectively. 
  
 The comparison between the isochrones and the observations reveals that a metallicity variation of $Z \sim 0.007$ is required to reproduce the large color difference observed between the red and the blue MS. However we note that the metal-rich and the metal-poor isochrones are well separated in the magnitude interval $21.00 \lesssim m_{\rm F814W}\lesssim 22.25$ in contrast with the observations, where the red and the blue MS merge together below $m_{\rm F814W} \sim 21.00$. 
Moreover, the magnitude difference between the MSTO of the isochrone with Z=0.151 and with Z=0.008 is about 0.2 mag in the F814W and is significantly smaller than the observed ones. These arguments prove that the split MS and the eMSTO of NGC\,1755 can not be explained in terms of metallicity variation.

The CMD of NGC\,1755 is clearly not consistent with distinct stellar populations with the same age and metallicity but different helium abundance as shown in the panel (b)  of Fig.~\ref{fig:isocrone2} where we compare the observed CMD with isochrones from the Padova Database with Y=0.26 (Girardi et al.\,2002, red line), Y=0.33 (green line), and Y=0.40 (blue line).  
 In panel (c) of  Fig.~\ref{fig:isocrone2} we further assume that the helium-rich isochrone (red line) is also more metal rich and younger than the isochrone with Y=0.26 (blue line). When this specific combination of age, metallicity, and helium is adopted, the MS of the helium-rich isochrone is redder than the MS with Y=0.26 for $ m_{\rm F814W}\lesssim 21$ while the two MSs merge together at fainter luminosity. However, the color separation between the two isochrones in the magnitude interval $m_{\rm F814W}\lesssim 21$ is significantly smaller than the corresponding color difference between the blue and the red MS. These facts demonstrate that helium variation alone as well as a combination of helium, metallicity, and age variation are not responsible for the split MS and the eMSTO of NGC\,1755.

%%%%
 The isochrones from the Geneva database (Ekstr{\"o}m et al.\,2012; Georgy et al.\,2013, 2014) with different rotation rates are superimposed on the observed CMD in panel (d) of Fig.~\ref{fig:isocrone2}. In this case, the blue isochrone has no rotation, while the green and the red isochrones have rotation, $\omega$, corresponding to 0.80 and 0.95 times the break-up value ($\omega_{\rm cr}$). The non-rotating isochrone has been extended to masses $\mathcal{M}<1.6\mathcal{M}_{\odot}$ by using the models of Mowlavi et al.\,(2012). 
 Unfortunately rotating isochrones for stars with $\mathcal{M}<1.6\mathcal{M}_{\odot}$ are not available to us. 
All the isochrones have the same metallicity, Z=0.006, and an age of 60 Myr. In order to match the observed CMD, we have adopted reddening E(B$-$V)=0.12 and distance modulus, (m$-$M)$_{0}$=18.38. 
 In this scenario the blue MS corresponds to the non-rotating isochrone, while the red MS is consistent with an isochrone with $\omega=0.95 \times \omega_{\rm cr}$.

\begin{centering} 
\begin{figure*} 
 \includegraphics[width=14.5cm]{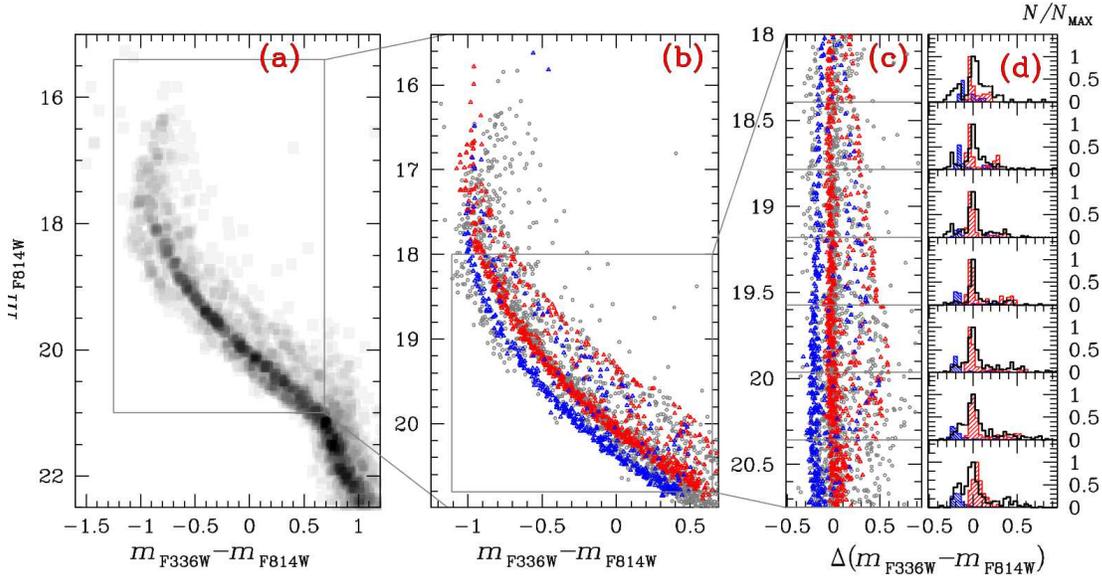} 
%/home/milone/NUBI/WFC3UVIS/NGC1755/figure/fig.macro go8
 \caption{ $m_{\rm F814W}$ vs.\,$m_{\rm F336W}-m_{\rm F814W}$ Hess diagram for stars in the NGC\,1755 field (panel a). Panel (b) compares the observed CMD (black dots) with the simulated CMD for non-rotating stars (blue symbols) and for rotating stars with $\omega=0.95 \times \omega_{\rm cr}$ (red symbols). The verticalized $m_{\rm F814W}$ vs.\,$\Delta$($m_{\rm F336W}-m_{\rm F814W}$) CMD around the region where the MS split is most evident is provided in panel (c). Panels (d) compare the distribution of the $\Delta$($m_{\rm F336W}-m_{\rm F814W}$) color in seven magnitude intervals for the observed stars (black histogram) and for the simulated  rotating and non-rotating stars (red- and blue-shaded histogram, respectively).} 
 \label{fig:SimRotazione} 
\end{figure*} 
\end{centering}

To further investigate  whether the split upper MS of NGC\,1755 can be explained with stellar populations having different rotation we adopted the procedure illustrated in Fig.~\ref{fig:SimRotazione}. In close analogy with what has been done by D'Antona et al.\,(2015) in their study of NGC\,1856, we have retrieved from the Geneva database the synthetic photometry corresponding to the blue and red isochrones used in the panel (d) of Fig.~\ref{fig:isocrone2}.
 They have assumed a random distribution for the viewing angle, adopted the gravity darkening model by Espinosa Lara \& Rieutord (2011) and included the limb-darkening effect (Claret 2000). We have used the model atmospheres by Castelli \& Kurucz\,(2013) to transform the synthetic simulations into the observational plane by adopting the transmission curves of the F336W and F814W filters of UVIS/WFC3. Moreover, we have added to the simulated photometry the photometric errors  determined in Sect.~\ref{sec:ms} and assumed a fraction of binaries, $f_{\rm bin}=0.41$ as derived in Sect.~\ref{sub:binaries}.

The simulated CMD is superimposed on the observed one in panel (b) of Fig.~\ref{fig:SimRotazione} while in panel (c) we show the corresponding verticalized  $m_{\rm F814W}$ vs.\,$\Delta$($m_{\rm F336W}-m_{\rm F814W}$) that has been derived as in Sect.~\ref{sub:pratio}. In both panels (b) %and panel (c) of Fig.~\ref{fig:SimRotazione} 
 we have used blue and red colors to represent non-rotating, and rotating stars respectively. In the panels (d) of  Fig.~\ref{fig:SimRotazione} we compare the observed $\Delta$($m_{\rm F336W}-m_{\rm F814W}$) distribution (black histograms) with the corresponding distribution of rotating and non-rotating stars in the simulated CMD (red- and blue-shaded histograms).
 As shown in Fig.~\ref{fig:SimRotazione}, the simulation based on Geneva models reproduces  very well the split upper MS of NGC\,1755 and supports the possibility that stellar rotation is the main responsible for the split MS and the eMSTO of NGC\,1755. 
  In this context, the evidence, shown in the right panel of Fig.~\ref{fig:reddening}, that the blue MS is more broadened in the pseudo color $C$ than the red MS could indicate that blue-MS stars span a range of rotation rates. As an alternative, the blue MS could host large fraction of non-interacting binaries with small mass ratio. 

%%%%%%%%%%%%%%%%%%%%%%%%%%%%%%%%%%%%%%%%%%%%%%%%%%%%%%%%%%%%%%%%%%%%%%%%%%
%%%%%%%%%%%%%%%%%%%%%%%%%%%%%%%%%%%%%%%%%%%%%%%%%%%%%%%%%%%%%%%%%%%%%%%%%%

\begin{centering} 
\begin{figure} 
 \includegraphics[width=9.0cm]{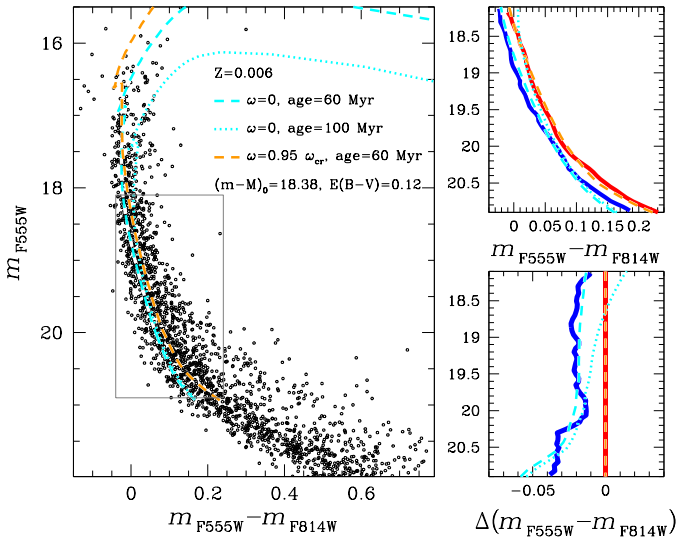} 
%/home/milone/NUBI/WFC3UVIS/NGC1755/figure/fig.macro test
 \caption{ We reproduce in the left panel the CMD obtained from WFC/ACS data and plotted in the central and right panel of Fig.~\ref{fig:ACS}. The orange and the cyan lines are the same isochrones used in the panel (d) of Fig.~\ref{fig:isocrone2} but in the $m_{\rm F555W}$ vs.\,$m_{\rm F555W}-m_{\rm F814W}$ plane, while the dotted line is a non-rotating 100-Myr old isochrone as indicated in the figure.  The CMD in the upper-right panel compares the isochrones with the fiducial line of the blue and the red MS (blue and red continuous line, respectively). The corresponding verticalized CMD is plotted in the lower-right panel.} 
 \label{fig:report1} 
\end{figure} 
\end{centering}

 In the right panel of Fig.~\ref{fig:report1} we compare the CMD $m_{\rm F555W}$ vs.\,$m_{\rm F555W}-m_{\rm F814W}$ with the same isochrones from Georgy et al.\,(2014) used in the panel (d) of Fig.~\ref{fig:isocrone2}. Specifically, the orange dashed line is the rotational isochrone with $\omega=0.95 \omega_{\rm cr}$ and the cyan dashed line is the non-rotational isochrone. We have assumed for all the isochrones shown in Fig.~\ref{fig:report1} the same values of metallicity, distance modulus, and reddening that we have previously determined from the analysis of the $m_{\rm F814W}$ vs.\,$m_{\rm F336W}-m_{\rm F814W}$ CMD.  The two isochrones represented with dashed lines have also the same age of 60 Myr as those shown in in the panel (d) of Fig.~\ref{fig:isocrone2}.  In addition we have shown a 100-Myr old non-rotating isochrone (cyan dotted line). 

 The comparison between the isochrones and the fiducial lines of the red and the blue MS is illustrated in the $m_{\rm F555W}$ vs.\,$m_{\rm F555W}-m_{\rm F814W}$ CMD  and in the verticalized $m_{\rm F555W}$ vs.\,$\Delta$($m_{\rm F555W}-m_{\rm F814W}$) CMD plotted in the upper-right and lower-right panels of Fig.~\ref{fig:report1}, respectively. 
The values of $\Delta$($m_{\rm F555W}-m_{\rm F814W}$) used in the verticalized diagram have been derived for the fiducial lines by subtracting from the $m_{\rm F555W}-m_{\rm F814W}$ color of each fiducial, the color of the red-MS fiducial at the corresponding  F555W magnitude. Similarly, we have determined $\Delta$($m_{\rm F555W}-m_{\rm F814W}$) for the isochrones by subtracting from the color of each isochrone the correspond color of the rotating isochrone.

Our analysis of the  $m_{\rm F555W}$ vs.\,$m_{\rm F555W}-m_{\rm F814W}$ CMD confirms that the MS split is well reproduced by two stellar populations with different rotation rates. In contrast, isochrones with different age are not able to properly reproduce the observed red and the blue MS.
 
 Noticeably, both the simulation shown in Fig.~\ref{fig:SimRotazione} and  the comparison between the CMD and the isochrones plotted in Fig.~\ref{fig:report1} are not fully satisfactory as the entire eMSTO and the SGB are poorly reproduced by the models. In priciple, we could improve the fit between the models and the observed MSTO by assuming that the non-rotating population is slightly older than the rotating ones. However, in this case, we would get a poor fit of the split MS as shown in the right panels of Fig.~\ref{fig:report1}. D'Antona et al.\,(2015) in their analysis of the $\sim$300 Myr old cluster NGC\,1856 have noticed a similar issue and suggested that the poor quality of the fit may be due to the fact that the end of the core hydrogen burning phase may be affected by several second-order parameters. These could include the way in which the inclination angle has been taken into account for the calculation of the stellar luminosity and effective temperature. 

 Moreover, in order to reproduce the split MS, we need that $\sim$75\% of stars are rapid rotators with $\omega=0.95 \times \omega_{\rm cr}$, that is very close to the critical value. 
As discussed by D'Antona et al.\,(2015), the large value of $\omega$ could be an artifact related to the specific treatment of gravity and limb darkening of these models. What is relevant here is that the adopted models from the Geneva group have allowed, for the first time, an appropriate representation of the split MS and the eMSTO in a young LMC cluster, thus providing strong indication that rotation is the main responsible for these CMD features.
%%%%%%%%%%%%%%%%%%%%%%%%%%%%%%%%%%%%%%%%%%%%%%%%%%%%%%%%%%%%%%%%%%%%%%%%%%
%%%%%%%%%%%%%%%%%%%%%%%%%%%%%%%%%%%%%%%%%%%%%%%%%%%%%%%%%%%%%%%%%%%%%%%%%%

 The SGB of intermediate-age GCs has been recently investigated to discriminate among the different scenarios proposed to interpret the CMD of these objects but results are still controversial. Li et al.\,(2014) have analyzed the CMD of NGC\,1651, whose eMSTO is consistent with an age spread of 450 Myr and concluded that its tight SGB excludes such a large age variation. Similar results have been obtained by Bastian \& Niederhofer (2015) from the analysis of NGC\,1806 and NGC\,1846.  In contrast, Goudfroij et al.\,(2015) based on recent stellar models that account for the effects of convective overshooting, have concluded that the color and magnitude distribution of SGB stars in NGC\,1651, NGC\,1806, and NGC\,1846 are consistent with those inferred from eMSTO when isochrones with different ages are adopted. As shown in Fig.~\ref{fig:SimRotazione}, the small number of SGB stars in both the observed and the simulated CMD prevents us from any conclusion based on the SGB in NGC\,1755. 

\section{Summary and Discussion}\label{sec:discussion} 
We have studied the LMC cluster NGC\,1755 by means of high-precision photometry obtained from the UVIS/WFC3 and ACS/WFC on board of {\it HST}.
 We have found that the MS of this $\sim$80-Myr old cluster is split into a blue and a red component hosting about the 25\% and 75\%, respectively, of the total number of MS stars in the analysed magnitude interval with $17.95<m_{\rm F814W}<20.75$. 
The double MS is visible over a range of more than three magnitudes in the $m_{\rm F814W}$ vs.\,$m_{\rm F336W}-m_{\rm F814W}$ CMD, from $m_{\rm F814W} \sim 17.75$ towards $m_{\rm F814W} \sim 21.0$ where the two MSs seem to merge together. The $m_{\rm F336W}-m_{\rm F814W}$ color separation is about 0.2 mag at the luminosity of the MSTO of the blue MS and slightly decreases towards the lower part of the MS. 
%Similarly, the two populations merge in the upper part of the MS for $m_{\rm F814W} \lessim 17.75$.  
  The TOs of the two MSs exhibit different luminosities with the MSTO of the red MS being approximatively 0.5 mag brighter than that of the red-MS turn off in the F814W band. These findings make NGC\,1755 the youngest cluster with eMSTO known to date, and,  together with NGC\,1844 and NGC\,1856 (Milone et al.\,2013, 2015), one of three Magellanic Cloud clusters with split MS.

We have demonstrated that the split MS and the eMSTO of NGC\,1755 is real and is not due to field-star contamination, photometric errors, or differential reddening. Moreover, we have determined the fraction of binary  systems and found that NGC\,1755 hosts a large fraction of binaries in the analysed field of view which is $f_{\rm bin}=0.41 \pm 0.03$. Non-interacting binaries are not responsible for the split MS.

In order to shed light on the physical mechanism responsible for the split upper MS we have compared the observed CMD with isochrones with different age, metallicity, and rotation rate and with synthetic CMDs.
  We have concluded that  the double MS and the eMSTO can not be explained in terms of neither variations in age nor metallicity variations. On the contrary, we show that the CMD is consistent with two stellar populations with different rotation.
%il caso di NGC1856
These results provide strong constraints on the main scenarios suggested to interpret the eMSTO phenomenon as we will discuss in the following subsections.

\subsection{Age difference}
%variazioni di eta' come spiegazione dei eMSTO
 The eMSTOs observed in most intermediate-age  Magellanic Cloud star clusters  have been interpreted by several groups as due to a extended star formation (e.g.\,Mackey \& Broby Nielsen 2007; Baume et al.\,2007; Glatt et al.\,2008; Milone et al.\,2009; Goudfrooij et al.\,2011). Moreover, Milone et al.\,(2015) and Correnti et al.\,(2015) have shown that the eMSTO of the $\sim$300\,Myr-old cluster NGC\,1856 is consistent with a prolonged period of star formation with a duration of about 150\,Myr.
 In contrast, the observed CMD of NGC\,1755 is not consistent with a difference in age. If we assume, that the same physical process is responsible  for both the eMSTO and the split MS of NGC\,1755, NGC\,1856, and for the eMSTO in the intermediate-age Magellanic Clouds star clusters, our findings rule out the possibility that these CMD features are mostly due to age difference.

Goudfrooij et al.\,(2014) have found a correlation between the MSTO width and the escape velocity of the host cluster and suggested that the eMSTO would occur in clusters with escape velocities greater than $\sim$15 km s$^{-1}$, whose velocity escape are higher than the velocity of the stellar winds of first-generation stars out of which the second generations form. Moreover Goudfrooij and collaborators have hypothesised an early-mass threshold of $\mathcal{M}_{\rm cl} \sim 10^{4.8} \mathcal{M}_{\odot}$ above which clusters seem to be able to host an eMSTO.

NGC\,1755 has a mass of $10^{4.0} \mathcal{M}_{\odot}$ (Popescu et al.\,2012) well below the threshold suggested by Goudfrooij et al.\,(2014).  
Thus,  if the same physical process is responsible for the eMSTO of young and intermediate-age Magellanic Clouds star clusters, the fact that the MSTO of this low-mass cluster is not consistent with a single isochrone would represent a challenge for the scenario by Goudfrooij and collaborators.  
%variazioni in metallicita'

\subsection{Metallicity variations}
Milone et al.\,(2015) have shown that, among other viable interpretations, the split MS and the eMSTO of NGC\,1856 could also be consistent with multiple stellar population with different metallicity.
 While variations in iron have been observed in several massive Galactic GCs (see Marino et al.\,2015  and references in their Table~10), high-precision photometry of intermediate-age star clusters have shown that the RGB are narrow and well defined and are not consistent with large metallicity variations.
 Moreover, Mucciarelli et al.\,(2014) have inferred the iron abundance of eight stars in the eMSTO cluster NGC\,1806 and have concluded that this cluster has homogeneous [Fe/H]. 
 The comparison between the observed CMD of NGC\,1755 and isochrones with different Z-values and ages rule out metallicity variation as responsible for the eMSTO phenomenon.

\subsection{Helium variations}
 The CMD of nearly all the old Galactic GCs is made of two or more sequences that can be followed along the entire CMD, from the RGB tip, to the hydrogen-burning limit, and correspond to stellar populations with different helium  content (Milone et al.\,2012c)  and light-element abundance (see Gratton et al.\,2012 for a review on the chemical composition of multiple stellar populations in GCs). The helium variation ranges from $\Delta$Y$\sim$0.15 as in $\omega$\,Centauri and NGC\,2808 (e.g.\,D'Antona et al.\,2002, 2005; \,Bedin et al.\,2004; Norris\,2004; Piotto et al.\,2007) to $\Delta$Y$\sim$0.01 as in NGC\,6397 and NGC\,288 (e.g.\,Milone et al.\,2012c; Piotto et al.\,2013; see Table~2 from Milone et al.\,2014).

In contrast, isochrones with different helium abundance do not reproduce the observed CMD of NGC\,1755. This fact provides a significant difference between the multiple sequences observed in the old Galactic GCs and in the young MC clusters.
\subsection{Interacting binaries}
 Merged binary systems and interacting binaries have been suggested as possible responsible for the eMSTO in intermediate-age clusters (Yang et al.\,2011). 
  Appropriate models, which unfortunately are not available to us, are mandatory to establish to what extend these binary systems are responsible for the eMSTO and the split MS of NGC\,1755.

  However, we note that the main challenge for this scenario is that, while the MSTO of intermediate-age MC clusters is either broadened or bimodal, their MS is narrow and well defined, thus suggesting that binaries would affect only a small mass interval of the CMD. 
 In addition,  binary systems and interactive binaries would produce a spread distribution of the color of stellar sources in the CMD and are not consistent with the split MS observed in NGC\,1755 and NGC\,1856.
 
\subsection{Rotation}
%variazioni in rotazione
 Multiple stellar populations with different rotation rates can mimic an age spread  in intermediate-age star clusters and can be responsible for their eMSTOs as first suggested by Bastian \& De Mink (2009). 
 Indeed the stellar structure is significantly affected by stellar rotation and its effective temperature depends on the angle between the star and the observer.

As discussed by Yang et al.\,(2013) the parameters adopted in the description of rotational mixing dramatically affect the rotational models so that the eMSTO of intermediate-age star clusters can be explained or not in terms of rotation depending on the adopted set of models.
 As an example the conclusion by Bastian \& De Mink (2009) is not supported by the results of Girardi et al.\,(2011) who have calculated isochrones for intermediate-age MC clusters and concluded that the eMSTO can not be reproduced by variations in the rotation rate. %Furthermore, the split MS of the young LMC NGC\,1844 is not properly reproduced by two stellar populations with different rotation when BaSTI models, modified as in Bastian \& De Mink (2009) to take into account the effects of rotation, are used (see Milone et al.\,2013 for details).
 More recently, Brand \& Huang\,(2015) and Niederhofer et al.\,(2015) have shown that rotating models by Georgy et al.\,(2014) can account for the eMSTO in intermediate-age clusters.

 Similar conclusion  that rotation is respondible for the eMSTO come from the study by D'Antona et al.\,(2015) of the $\sim$300-Myr old LMC cluster NGC\,1856. 
 This cluster exhibits both an eMSTO and a split MS with the blue MS hosting about one third of the total number of MS stars. 
 D'Antona and collaborators have shown that the CMD of NGC\,1856 is consistent with two stellar populations. A group of non-rotating or slow-rotating stars that correspond to the blue MS and the faint MSTO, plus a population of fast rotating stars with $\omega=0.90 \times \omega_{\rm cr}$ that are associated to the red MS and the bright MSTO. As in the present case, the authors warn that the choice of $\omega=0.90 \times \omega_{\rm cr}$  may simply represent a formal solution, linked to the choices of Georgy et al.\,(2014)  concerning limb and gravity darkening.

In this paper, we have used models from  Georgy et al.\,(2014) and find that the observed CMD of NGC\,1755 is consistent with two stellar populations with different rotation rates. A stellar population hosting about one-forth of the total number of stars with no rotation plus a population of fast rotating stars with $\omega=0.95 \times \omega_{\rm cr}$. 
 While the CMD of NGC\,1856 was also consistent with age or metallicity variations (Milone et al.\,2015; Correnti et al.\,2015), the observations of the young NGC\,1755 rule out age and metallicity variations as responsible for the eMSTO and the split MS.
 Unfortunately, rotating models for stars with $\mathcal{M}<1.6\mathcal{M}_{\odot}$ are not available to us, and our conclusions are based on the comparison between the models and the CMD region with $m_{\rm F814W} \lesssim 21$. 

 Noticeably, the interpretation of the eMSTO phenomenon in MC clusters is due to stellar populations with different rotation is still controversial.
 First of all, multiple MSs have been detected only in three clusters younger than $\sim$300 Myr, namely NGC\,1844, NGC\,1755 and NGC\,1856 and the eMSTO has been observed only in the latter two clusters. Appropriate observations of a large sample of young clusters are needed to establish to what extent the multiple MSs and eMSTO are common features of these objects. Moreover, there are significant difference in the interpretation of the eMSTO in intermediate-age and young clusters.
 As an example, the conclusion by Niederhofer et al.\,(2015) that the eMSTO of intermediate-age clusters can be explained in terms of stellar rotation, are valid as long as the initial rotation distribution covers the interval from $\omega \sim 0.0$ and $\omega \sim 0.5\omega_{\rm cr}$. In contrast, the split MS  of NGC\,1755 and NGC\,1856 are consistent with two distinct populations with extreme values of $\omega=0.0$ and $\omega \sim 0.9 \times \omega_{\rm cr}$.  We also emphasize that the rotational models  adopted in this paper provide a poor fit of the eMSTO of NGC\,1755 although this issue could by due to second-order parameters that affect the hydrogen burning phase.  As an example, the stellar luminosity and effective temperature are significantly affected by the way the in which the inclination angle has been taken into account (see D'Antona et al.\,2015 for details).

%%%%%%%%%%%
As previously noticed by D'Antona et al.\,(2015) for NGC\,1856, the origin of the non-rotating or slow-rotating stellar populations is another open issue and a possible solution could come from studies on B- and A-type stars in the field. Since most of the field B- and A-type stars are rapid rotators, the main challenge would concern the physical process responsible for slowing down about one forth of stars in NGC\,1755.
 A bimodal velocity distribution, with a slow-velocity component, has been observed  by Zorec \& Royer (2012) among MS field stars with $2.40 \mathcal{M}_{\odot}< \mathcal{M}< 3.85\mathcal{M}_{\odot}$ and by Dufton et al.\,(2013) among early B-type stars.
This analysis was mostly appropriate for NGC\,1856 stars, with turnoff masses of $\sim 3\mathcal{M}_{\odot}$. Here we are dealing also with masses considerably larger, up to 5 $\mathcal{M}_{\odot}$ or more, for which the observational data are scarce. It is well possible that we are witnessing the same kind of process.

 Zorec \& Royer (2012) have suggested that the observed slow-rotating and non-rotating stars would have lost their angular momentum during the pre-MS phase due to magnetic breaking. As an alternative, tidal interaction in binary systems could be responsible for the slow down of a fraction of stars.

 The latter hypothesis is supported by the study of B- and A-type stars by Abt \& Boonyarak (2004). These authors have found that, due to tidal interaction, the rotational velocities in binaries with period between 4 and 500 days is significantly smaller than for single stars.
D'Antona et al.\,(2015) have thus suggested that binary synchronisation could be responsible for the slow-rotating population of NGC\,1856 and we can extend this view to the case of NGC\,1755. 

In this context, we note that the fraction of non-rotating stars in the $\sim$300 Myr old cluster NGC\,1856 is the 33$\pm$5\% of the total number of MS stars (Milone et al.\,2015) and is slightly higher than that observed in the younger cluster NGC\,1755. These values are similar to the fraction of stars in the faint MSTO of the intermediate-age clusters NGC\,1806 (26$\pm$4\%), NGC\,1846 (25$\pm$3\%), and NGC\,1751 (31$\pm$4\%, Milone et al.\,2009) and suggest that, whatever the mechanism for slowing down rotation, it must be a fast process, so that there is no correlation between the cluster age and the fraction of non-rotating stars. Otherwhise, we must assume that a fraction of $\sim$25-30\% of cluster stars are born slowly rotating.

\section*{acknowledgments} 
\small 
APM, AFM, GDC, HJ, and DM acknowledge support by the Australian Research Council through Discovery Early Career Researcher Awards DE150101816 and DE160100851 and Discovery Project grants  DP150103294 and DP150100862. FD acknowledges support from PRIN-INAF 2014 (P.I.\,S.\,Cassisi). We thank the anonymous referee for several suggestions that have improved the quality of the manuscript.

\bibliographystyle{aa}

\end{document}